\documentclass[aps,preprint,amsmath,amssymb,amsfonts]{revtex4}
\usepackage{epsfig}
\usepackage{graphicx}
\usepackage{subfigure}
\usepackage{dcolumn}
\usepackage{bm}
\usepackage{amsthm}
\usepackage{enumitem}
\usepackage{slashed}
\usepackage{braket}
\usepackage{amsmath}
\usepackage{mathtools}

\makeatletter
\def\l@subsubsection#1#2{}
\makeatother

\newcommand{\del}{\partial}

\renewcommand{\Im}{\operatorname{Im}}
\newcommand{\sign}{\operatorname{sign}}

\newcommand{\Res}{\operatorname{Res}}

\newcommand{\bbR}{\mathbb{R}}

\newcommand{\calA}{\mathcal{A}}

\newcommand{\calF}{\mathcal{F}}
\newcommand{\calG}{\mathcal{G}}

\newcommand{\calJ}{\mathcal{J}}

\newcommand{\calM}{\mathcal{M}}
\newcommand{\calN}{\mathcal{N}}

\newcommand{\calS}{\mathcal{S}}

\newcommand{\scri}{\mathcal{I}}

\begin{document}

\title{MHV amplitudes and BCFW recursion for Yang-Mills theory in the de Sitter static patch}

\author{Emil Albrychiewicz}
\email{ealbrych@berkeley.edu}
\affiliation{University of California, Berkeley, CA 94720, U.S.A.}
\author{Yasha Neiman}
\email{yashula@icloud.com}
\affiliation{Okinawa Institute of Science and Technology, 1919-1 Tancha, Onna-son, Okinawa 904-0495, Japan}
\author{Mirian Tsulaia}
\email{mirian.tsulaia@oist.jp}
\affiliation{Okinawa Institute of Science and Technology, 1919-1 Tancha, Onna-son, Okinawa 904-0495, Japan}

\date{\today}

\begin{abstract}
We study the scattering problem in the static patch of de Sitter space, i.e. the problem of field evolution between the past and future horizons of a de Sitter observer. We formulate the problem in terms of off-shell fields in Poincare coordinates. This is especially convenient for conformal theories, where the static patch can be viewed as a flat causal diamond, with one tip at the origin and the other at timelike infinity. As an important example, we consider Yang-Mills theory at tree level. We find that static-patch scattering for Yang-Mills is subject to BCFW-like recursion relations. These can reduce any static-patch amplitude to one with N\textsuperscript{-1}MHV helicity structure, dressed by ordinary Minkowski amplitudes. We derive all the N\textsuperscript{-1}MHV static-patch amplitudes from self-dual Yang-Mills field solutions. Using the recursion relations, we then derive from these an infinite set of MHV amplitudes, with arbitrary number of external legs.
\end{abstract}

\maketitle
\tableofcontents
\newpage

\section{Introduction} \label{sec:intro}

\subsection{Scattering in finite regions of Minkowski and de Sitter space}

In the last decades, theoretical physics got progressively better at calculating observables defined on the boundary of spacetime. These include scattering amplitudes in Minkowski space, as well as correlators at conformal infinity of dS or AdS. There are good reasons for focusing on such observables. First, they're relatively easy to calculate. Second, they're observationally relevant: the flat S-matrix describes collider experiments, whereas correlations at dS future infinity encode the consequences of the conjectured inflationary epoch \cite{Maldacena:2002vr}. Finally, in quantum gravity, observables at infinity are the \emph{only} ones we know how to make sense of (with the AdS case the best-understood, via AdS/CFT \cite{Maldacena:1997re,Gubser:1998bc,Witten:1998qj,Aharony:1999ti}).

For all these reasons, the evolution of systems confined to finite regions of space has received less attention within fundamental theory. As a simple example, consider a spherical region in flat spacetime. The causal development of such a region is a causal diamond, bounded by the lightcone of one point in the past, and the lightcone of another point in the future. One can then define a ``scattering'' problem for this finite region: to calculate the fields (or the quantum state) on the final lightcone in terms of those on the initial lightcone. Almost no work on this problem exists. 

Is this just the proper state of affairs? Should we dismiss such finite-region observables as some combination of complicated, pointless and (in quantum gravity) ill-defined? Tempting though this may be, we have an observational fact to contend with -- the accelerated expansion of our Universe, which appears consistent with a positive cosmological constant, and thus a de Sitter asymptotic future. This implies a cosmological horizon that asymptotes to a finite size, with no access -- even in principle -- to spatial infinity. De Sitter space does have a \emph{future} conformal boundary, but that only becomes observable if the accelerated expansion eventually ends, as in inflation. If we believe that we are \emph{truly} stuck in an asymptotically dS world, we must come to term with physics without observables at infinity. For quantum gravity, this is a tall order indeed. However, we can take baby steps, by familiarizing ourselves with field-theory questions that are natural for an observer inside a de Sitter cosmological horizon. 

For simplicity, we now leave real-world cosmology aside, and consider \emph{pure} de Sitter space $dS_4$ -- the simplest spacetime in which every observer is trapped inside a spherical horizon of finite size. The largest spacetime region available to such an observer is a \emph{static patch} of $dS_4$; see discussions in e.g. \cite{Anninos:2011af,Halpern:2015zia}. The static patch is bounded by a past horizon and a future horizon -- the lightcones of the past and future endpoints of the observer's worldline. Confined to such a region, the closest thing we have to an asymptotic observable is the ``S-matrix'' encoding field evolution from the past horizon to the future one. This general problem -- the problem of static-patch scattering -- was posed and studied by us in \cite{David:2019mos,Albrychiewicz:2020ruh}. One of our long-term goals is to work out this scattering problem within the context of higher-spin gravity \cite{Vasiliev:1995dn,Vasiliev:1999ba} -- a gravity-like theory of massless interacting fields with all spins, whose holographic description \cite{Klebanov:2002ja,Sezgin:2002rt,Sezgin:2003pt,Giombi:2012ms} appears to carry over from AdS to dS \cite{Anninos:2011ui}, and which apparently can be formulated on a pure, non-fluctuating $dS_4$ geometry \cite{Neiman:2015wma}.

On the route to higher-spin theory, the more ordinary theories of interacting massless fields form natural stepping stones. We studied free massless fields of all spins in \cite{David:2019mos}, and a scalar with cubic interaction in \cite{Albrychiewicz:2020ruh}. The natural next case is interacting spin-1, i.e. Yang-Mills theory. This will be our subject in the present paper. We will restrict our analysis to tree-level, where YM theory enjoys the simplifying property of conformal symmetry. As a result, it doesn't actually see the curvature of de Sitter space. All that remains of the de Sitter static patch is its conformal structure, which is identical to that of a causal diamond in Minkowski (the two are also conformal to the Rindler wedge of Minkowski, and to the static hyperbolic space $\bbR\times H_3$). Thus, in this case, the cosmologically motivated problem of static-patch scattering is actually equivalent to the flat causal-diamond problem we mentioned before. 

We are thus pursuing two goals. The first is just to study finite-region scattering, using the simple case of a conformal theory in a (conformally) flat causal diamond. The second is to study the de Sitter static patch specifically, in the hope that the simple case of YM theory will provide a useful stepping stone towards perturbative GR, and ultimately higher-spin gravity.

\subsection{Outline of the paper}

As always, it is crucial to set up the calculation in a way that makes best use of available symmetries. A priori, the Minkowski causal diamond is quite challenging in this regard, since its only isometries are $SO(3)$ rotations. The $dS_4$ static patch is only slightly better, with a symmetry of $\bbR\times SO(3)$, where the $\bbR$ describes time translations. In \cite{Albrychiewicz:2020ruh}, we proposed a general strategy for the static-patch problem, which makes use of the larger symmetry of $dS_4$ as a whole. This involves artificially extending the static patch's boundaries into geodesically complete cosmological horizons, each one defining a Poincare patch. These are endowed with spatial translation symmetry, making it possible to work in momentum space. The static-patch problem was then decomposed into a pair of Poincare-patch evolutions, sewn together by a coordinate inversion at the conformal boundary of $dS_4$. Under such an inversion, the spatial translations of one Poincare patch become the conformal boosts of the other, and vice versa.

In the present paper, we take a simpler approach, taking advantage of the fact that we are dealing with a conformal theory. The \emph{conformal} symmetry of a Minkowski causal diamond, or a $dS_4$ static patch, is $\bbR\times SO(1,3)$. This is better than the non-conformal case, but not good enough on its own. Just as in \cite{Albrychiewicz:2020ruh}, we can gain \emph{translation symmetry} (in this case, full 4d Minkowski translations) by artificially extending our scope outside the causal diamond. The first step is to choose a flat conformal frame in which the causal diamond's tips are fixed at (past timelike) infinity and the origin. In this frame, the diamond's past boundary (or the past horizon of the de Sitter observer) becomes a portion of past null infinity $\scri^-$, while its future boundary (the de Sitter observer's future horizon) becomes the origin's past lightcone. We can then artificially extend the initial data to \emph{all of} $\scri^-$, and decompose it into plane waves. As we will see in more detail below, this reduces our scattering problem to a calculation of bulk YM fields in a lightcone gauge, out of plane-wave initial data. In this setup, the analog of an $n$-point scattering amplitude is encoded in the bulk field at $(n-1)$'st order in the initial data. This quantity is sometimes called a Berends-Giele current, after the work \cite{Berends:1987me}. Thus, the static-patch problem is reduced to a fairly standard Minkowski calculation, involving $n-1$ ingoing on-shell legs, and one outgoing off-shell leg (there is no loss of generality in having \emph{one} outgoing leg, since we'll be working at the level of field operators rather than Fock states).

Having thus framed the scattering problem, we will proceed to tackle it at tree level. Our main results are as follows:
\begin{enumerate}
	\item The simplest non-vanishing static-patch amplitudes are those with N\textsuperscript{-1}MHV helicity structure, i.e. those in which all but one external leg have the same helicity. These are closely related to the Parke-Taylor MHV amplitudes \cite{Parke:1986gb} of the Minkowski S-matrix. We will derive these N\textsuperscript{-1}MHV static-patch amplitudes from perturbative self-dual solutions of the Yang-Mills equations \cite{Bardeen:1995gk,Cangemi:1996rx,Korepin:1996mm,Rosly:1996vr}, closely following \cite{Rosly:1996vr}.  
	\item All other static-patch amplitudes can be reduced to these N\textsuperscript{-1}MHV ones, using an appropriately modified version of the BCFW recursion relations \cite{Britto:2004ap,Britto:2005fq}. This constitutes a modern upgrade over Berends and Giele's original recursion relations \cite{Berends:1987me} for the Berends-Giele current: while those relations basically translate into a sum over all Feynman diagrams, our BCFW-type relations will generally involve fewer summands.
\end{enumerate}
We emphasize that this is more than what's been achieved for standard (A)dS boundary correlators. Just like our static-patch problem, the (A)dS boundary problem for tree-level Yang-Mills can be conformally transformed into flat spacetime, where the (A)dS boundary becomes just a flat hypersurface $z=0$. And yet, this problem is not so easy! In particular, already N\textsuperscript{$-2$}MHV correlators are non-zero, and already for them there isn't a known formula for general $n$. Correlator formulas in spinor-helicity language are only known for $n=3$ \cite{Maldacena:2011nz} and $n=4$ \cite{Armstrong:2020woi} (see also \cite{Nagaraj:2018nxq,Nagaraj:2019zmk,Nagaraj:2020sji}), with results \cite{Albayrak:2018tam} outside the spinor-helicity formalism for $n=5,6$. Some recursion relations have also been developed in this context \cite{Raju:2012zr,Raju:2012zs,Albayrak:2019asr}, but these either leave a $z$ integration to be performed \cite{Raju:2012zr,Raju:2012zs}, or else pertain only to one Witten diagram at a time \cite{Albayrak:2019asr}. Thus, for Yang-Mills theory, the static-patch scattering problem, while less trivial than the Minkowski S-matrix, is \emph{easier} than standard (A)dS boundary correlators, despite having nominally lower symmetry. The simplification can be ascribed to the lightlike nature of the static patch's boundary. This allows helicity to be represented naturally in terms of ``little-group'' $SO(2)$ rotations around the lightrays, as opposed to the $SO(3)$ rotational structure on the non-lightlike conformal boundary of (A)dS. As for the static patch's lower overall symmetry, it is in large part compensated by the fact that the static-patch problem has a square root, as in \cite{Albrychiewicz:2020ruh}: the past and future horizons can be regarded separately, which gives us access to the symmetry of the Poincare patch -- or, in our case, due to YM theory being tree-level conformal, to the full symmetry of Minkowski space.

The rest of the paper is structured as follows. In section \ref{sec:kinematics}, we define the static-patch scattering problem natively in de Sitter space, and then gradually transform it into a more-or-less standard calculation in Minkowski, in a spinor-helicity formalism. The final results of that section are given in eqs. \eqref{eq:kinematics}-\eqref{eq:amplitudes_from_fields}. In section \ref{sec:pre_MHV}, we derive all the tree-level N\textsuperscript{-1}MHV static-patch amplitudes from a self-dual Yang-Mills solution and an anti-self-dual perturbation over it. The results for these amplitudes are given in eqs. \eqref{eq:pre_MHV_from_self_dual},\eqref{eq:pre_MHV_from_self_dual_opposite},\eqref{eq:pre_MHV_from_f_left},\eqref{eq:pre_MHV_from_f_left_opposite}. In section \ref{sec:poles}, we discuss the pole structure of tree-level static-patch amplitudes, relating them to the standard Minkowski S-matrix, and proving a BCFW-type recursion relation. The final form of this recursion relation is given in eqs. \eqref{eq:BCFW}-\eqref{eq:BCFW_opposite}. In section \ref{sec:MHV}, we apply the recursion to compute a class of MHV static-patch amplitudes. These are given in eqs. \eqref{eq:MHV_result}-\eqref{eq:MHV_result_opposite}. Section \ref{sec:discuss} is devoted to discussion and outlook.

\section{Geometry and kinematics} \label{sec:kinematics}

In this section, we set up the geometry and kinematics of the static-patch problem and its Minkowski reformulation. In section \ref{sec:kinematics:static}, we describe the $dS_4$ static patch, and define the initial and final field data on its past and future horizons. In section \ref{sec:kinematics:flat}, we introduce Poincare coordinates adapted to the past horizon, and the associated conformal transformation into Minkowski space. In section \ref{sec:kinematics:linear}, we construct linearized Yang-Mills solutions in Minkowski, and relate them to initial data on the past horizon. Finally, in section \ref{sec:kinematics:non_linear}, we discuss non-linear bulk fields, and show how certain components of them in a certain gauge correspond to final data on the future horizon.

\subsection{The static-patch problem, formulated in embedding space} \label{sec:kinematics:static}

De Sitter space $dS_4$ is the hyperboloid of unit spacelike radius inside the flat 5d embedding space $\bbR^{1,4}$. We use lightcone coordinates $r^I = (u,v,\mathbf{r})$ for $\bbR^{1,4}$, where boldface indicates a 3d Euclidean vector. The $\bbR^{1,4}$ metric reads:
\begin{align}
 ds^2 = -2dudv + \mathbf{dr}^2 \ , \label{eq:dS_metric}
\end{align}
and the $dS_4$ hyperboloid is given by:
\begin{align}
 -2uv + \mathbf{r}^2 = 1 \ . \label{eq:dS}
\end{align} 
The curved metric of $dS_4$ is just the flat 5d metric \eqref{eq:dS_metric}, restricted to the hyperboloid \eqref{eq:dS}.

The $(u,v,\mathbf{r})$ coordinate system is adapted to a particular observer in $dS_4$ -- the one whose worldline begins at $(u,v)=(0,-\infty)$ and ends at $(u,v)=(\infty,0)$. The observer's past horizon is given by $(u=0,v<0,\mathbf{r}^2=1)$, and her future horizon is given by $(u>0,v=0,\mathbf{r}^2=1)$. Each horizon is a lightlike cylinder, consisting of a spatial unit sphere $\mathbf{r}^2=1$, multiplied by the lightlike $u$ or $v$ axis. See the Penrose diagram in Figure \ref{fig:Penrose}(a).

We will be dealing with a YM field on $dS_4$, which can be written as a 1-form $\hat A_I = (\hat A_u,\hat A_v,\mathbf{\hat A})$. The hats are to distinguish these components in the $(u,v,\mathbf{r})$ basis from the ones we'll introduce in a Poincare-patch basis below. We set to zero the component $\hat A_I r^I$ that points outside the $dS_4$. We will not explicitly write color indices; instead, we understand $\hat A_I$ to take values in the gauge algebra. While final answers can only depend on the Lie bracket, i.e. on the commutators of gauge algebra elements, it will be very convenient to work as if they have an associative product. This can be made concrete by defining the $\hat A_I$ as matrices over the gauge group's fundamental representation. We will not make assumptions about the gauge group itself and its structure constants, and we'll simply keep different product orderings as distinct terms. This attitude, taken from \cite{Rosly:1996vr}, is analogous to the now standard decomposition of the YM S-matrix into separately considered color-ordered pieces. 

Our task will be to express the YM field on the future horizon in terms of that on the past horizon. As discussed in \cite{Albrychiewicz:2020ruh}, this is a bit different from the usual scattering problem in Minkowski, where we are interested in the S-matrix relating Fock states on $\scri^-$ to Fock states on $\scri^+$. We will consider this distinction in more detail in section \ref{sec:kinematics:flat:fields_vs_states}.

Now, what should we take as the initial (final) field data on the past (future) horizon? First of all, since the horizons are lightlike, it is sufficient to consider the value of $\hat A_I$ on them: there is no need to separately include the normal derivative. Second, due to gauge freedom in the $dS_4$ bulk, it is sufficient to consider the components of $\hat A_I$ \emph{along} each horizon, i.e. $(\hat A_v,\mathbf{\hat A}_\perp)$ on the past horizon and $(\hat A_u,\mathbf{\hat A}_\perp)$ on the future one, where $\mathbf{\hat A}_\perp \equiv \mathbf{\hat A - (\hat A\cdot r)r}$ denotes the components along the 2-sphere $\mathbf{r}^2 = 1$. Finally, we fix the residual gauge freedom on each horizon by taking the potential along its lightrays to vanish. This sets $\hat A_v=0$ ($\hat A_u=0$) on the past (future) horizon respectively, leaving just the spatial components $\mathbf{\hat A}_\perp$ along the 2-sphere. Equivalently, we can work with the derivative of $\mathbf{\hat A}_\perp$ along each horizon's lightrays, which in our chosen gauge encodes the field strength components $\mathbf{\hat F}_{v\perp} = \del_v\mathbf{\hat A}_\perp$ or $\mathbf{\hat F}_{u\perp} = \del_u\mathbf{\hat A}_\perp$. In Maxwell theory, we could forget about gauge altogether and just focus on these field strength components. However, in YM theory, the field strength isn't gauge-invariant, and so the gauge choice $\hat A_v=0$ or $\hat A_u=0$ remains important.

So far, then, our initial data on the past horizon is given by $\mathbf{\hat A}_\perp(0,v,\mathbf{r})$, and the final data on the future horizon -- by $\mathbf{\hat A}_\perp(u,0,\mathbf{r})$. For our scattering calculation, we will want to express the initial data as plane waves in the Poincare coordinates associated with the past horizon. As we will see below, this simply requires a Fourier transform with respect to $v$ \cite{David:2019mos,Albrychiewicz:2020ruh}. In the interest of treating both horizons symmetrically, we'll Fourier-transform the final data with respect to $u$ as well. Now, recall that the boundaries of our static patch are actually ``half-horizons'', confined to the lightlike coordinate ranges $v<0$ and $u>0$; the other half of each horizon is unobservable. When we Fourier-transform the initial (final) data with respect to $v$ ($u$), we actually extend the original scattering problem. In the extended problem, we calculate the final fields on the \emph{entire} horizon $(u\in\bbR,v=0)$ -- a lightlike initial data hypersurface for the entire $dS_4$ spacetime, as a functional of the initial fields on the entire horizon $(u=0,v\in\bbR)$ -- another such hypersurface. This extension of the problem, which will prove convenient, is quite harmless. Indeed, at the very end, we can always limit our attention to the observable final fields at $u>0$. Since our field theory in $dS_4$ is causal, these can only depend on the initial fields in the observable range $v<0$; this dependence then defines our sought-after static-patch scattering \footnote{Meanwhile, the data on the unobservable future half $v>0$ of the past horizon will evolve into data on the unobservable past half $u<0$ of the future horizon. This describes scattering in the opposite static patch, with an opposite time orientation (since the $(u=0,v>0)$ half-horizon is actually to the \emph{future} of the $(u<0,v=0)$ half-horizon).}.

With this understood, we proceed to package the initial data into (gauge-algebra-valued) Fourier coefficients, as:
\begin{align}
 \mathbf{c}_{\text{in}}(\omega,\mathbf{r}) = \int_{-\infty}^\infty dv\,\mathbf{\hat A}_\perp(0,v,\mathbf{r})\,e^{i\omega v} \ . \label{eq:vector_modes_initial}
\end{align}
As we will see, in the Poincare coordinates associated with the past horizon, the coefficients $\mathbf{c}_{\text{in}}(\omega,\mathbf{r})$ describe lightlike plane waves, with 4-momentum:
\begin{align}
 k^\mu = (\omega,\mathbf{k}) = (\omega,-\omega\mathbf{r}) \ , \label{eq:k}
\end{align}
where the minus sign stems from the fact that a wave traveling along $\mathbf{k}$ is \emph{coming from} the direction of $-\mathbf{k}$ at past null infinity.

We can now introduce spinor-helicity variables, by taking the spinor square root of this momentum \cite{Maldacena:2011nz,David:2019mos}. We begin by introducing $SO(3)$ spinors $\psi^\alpha$, whose indices are raised and lowered as $\psi_\alpha = \epsilon_{\alpha\beta}\psi^\beta$ and $\psi^\alpha = \psi_\beta\epsilon^{\beta\alpha}$, with spinor complex conjugation acting as $\bar\psi_\alpha = (\psi^\alpha)^*$, and with the Pauli matrices $\boldsymbol{\sigma}^\alpha{}_\beta$. Now, for each 4-momentum of the form \eqref{eq:k}, we define its spinor square root $(\lambda^\alpha,\tilde\lambda_\alpha)$ via:
\begin{align}
  \langle\tilde\lambda\lambda\rangle \equiv \tilde\lambda_\alpha\lambda^\alpha = 2\omega \ ; \quad \langle\tilde\lambda\boldsymbol{\sigma}\lambda\rangle \equiv \tilde\lambda_\alpha\boldsymbol{\sigma}^\alpha{}_\beta\lambda^\beta = 2\mathbf{k} \ , \label{eq:spinor_helicity_3d}
\end{align}
where the factors of 2 are for later convenience. The reality of $k^\mu$ implies that the spinors $(\lambda^\alpha,\tilde\lambda_\alpha)$ are related by complex conjugation, up to the sign of the energy $\omega$:
\begin{align}
 \tilde\lambda_\alpha = \sign(\omega)\bar\lambda_\alpha \ .
\end{align}
Note also that, as usual, \eqref{eq:spinor_helicity_3d} defines $(\lambda^\alpha,\tilde\lambda_\alpha)$ only up to multiplication by opposite complex phases:
\begin{align}
 \lambda^\alpha \rightarrow e^{i\phi}\lambda^\alpha \ ; \quad \tilde\lambda_\alpha \rightarrow e^{-i\phi}\tilde\lambda_\alpha \ . \label{eq:phase_rotation}
\end{align}
One advantage of spinor-helicity variables is that the polarizations of $\mathbf{\hat A}_\perp$ can be decomposed into two helicities, given by the null complex vectors $\langle\lambda\boldsymbol{\sigma}\lambda\rangle$ and $\langle\tilde\lambda\boldsymbol{\sigma}\tilde\lambda\rangle = \langle\bar\lambda\boldsymbol{\sigma}\bar\lambda\rangle$. More precisely, we will use the following normalized versions of these vectors:
\begin{align}
 \mathbf{m} =  -\frac{\langle\lambda\boldsymbol{\sigma}\lambda\rangle}{\langle\tilde\lambda\lambda\rangle} \ ; \quad 
 \mathbf{\bar m} = \frac{\langle\tilde\lambda\boldsymbol{\sigma}\tilde\lambda\rangle}{\langle\tilde\lambda\lambda\rangle} \ ; \quad \mathbf{m\cdot\bar m} = 2 \ . \label{eq:m}
\end{align}
Extracting the components of \eqref{eq:vector_modes_initial} along $\mathbf{m}$ and $\mathbf{\bar m}$, we obtain the initial mode coefficients as spinor-helicity functions:
\begin{align}
\begin{split}
  c^+_{\text{in}}(\lambda^\alpha,\tilde\lambda_\alpha) &= -\frac{\langle\lambda\boldsymbol{\sigma}\lambda\rangle}{\langle\tilde\lambda\lambda\rangle}\cdot
    \int_{-\infty}^\infty dv\,\mathbf{\hat A}\!\left(0,v,-\frac{\langle\tilde\lambda\boldsymbol{\sigma}\lambda\rangle}{\langle\tilde\lambda\lambda\rangle}\right) e^{i\langle\tilde\lambda\lambda\rangle v/2} \ ; \\
  c^-_{\text{in}}(\lambda^\alpha,\tilde\lambda_\alpha) &= +\frac{\langle\tilde\lambda\boldsymbol{\sigma}\tilde\lambda\rangle}{\langle\tilde\lambda\lambda\rangle}\cdot
    \int_{-\infty}^\infty dv\,\mathbf{\hat A}\!\left(0,v,-\frac{\langle\tilde\lambda\boldsymbol{\sigma}\lambda\rangle}{\langle\tilde\lambda\lambda\rangle}\right) e^{i\langle\tilde\lambda\lambda\rangle v/2} \ ,
 \end{split} \label{eq:spinor_modes_initial}
\end{align}
where the superscript $\pm$ denotes helicity. Under the phase rotation \eqref{eq:phase_rotation}, the coefficients $c^\pm_{\text{in}}$ transform with weight $\pm 2$ respectively. The reality condition on the fields (which we will not impose) is $c^-_{\text{in}}(\lambda,\tilde\lambda) = -\overline{c^+_{\text{in}}(\lambda,-\tilde\lambda)}$. 

For the final data on the future horizon, we define spinor-helicity functions $c^\pm_{\text{out}}(\mu^\alpha,\tilde\mu_\alpha)$ in complete analogy with \eqref{eq:spinor_modes_initial}:
\begin{align}
 \begin{split}
   c^+_{\text{out}}(\mu^\alpha,\tilde\mu_\alpha) &= -\frac{\langle\mu\boldsymbol{\sigma}\mu\rangle}{\langle\tilde\mu\mu\rangle}\cdot\int_{-\infty}^\infty du\,
      \mathbf{\hat A}\!\left(u,0,-\frac{\langle\tilde\mu\boldsymbol{\sigma}\mu\rangle}{\langle\tilde\mu\mu\rangle}\right) e^{i\langle\tilde\mu\mu\rangle u/2} \ ; \\
   c^-_{\text{out}}(\mu^\alpha,\tilde\mu_\alpha) &= +\frac{\langle\tilde\mu\boldsymbol{\sigma}\tilde\mu\rangle}{\langle\tilde\mu\mu\rangle}\cdot\int_{-\infty}^\infty du\,
      \mathbf{\hat A}\!\left(u,0,-\frac{\langle\tilde\mu\boldsymbol{\sigma}\mu\rangle}{\langle\tilde\mu\mu\rangle}\right) e^{i\langle\tilde\mu\mu\rangle u/2} \ .
 \end{split} \label{eq:spinor_modes_final}
\end{align}
Again, though we are ultimately interested in the observable half-horizon $u>0$, it will be more convenient to work with Fourier coefficients on the entire $u$ axis, as in \eqref{eq:spinor_modes_final}. At the very end, we can restrict attention to the original static patch by simply throwing away the unobservable $u<0$ portion of the final data. Just like the initial modes \eqref{eq:spinor_modes_initial}, the final modes \eqref{eq:spinor_modes_final} also describe lightlike plane waves, but in a different Poincare frame -- the one adapted to the future horizon. In \cite{Albrychiewicz:2020ruh}, we explicitly made use of both Poincare frames. However, in our present case of a conformal theory, it will be simpler to just stay in one of them -- for concreteness, the one adapted to the past horizon \footnote{After submitting the first version of this manuscript, we understood that using one Poincare frame is easier for general massless theories, even with non-conformal interactions. This will be demonstrated in a future publication.}. There is a price to be paid: as we'll see, in this approach, the gauge choice $\hat A_u=0$ on the final horizon is easy to enforce only on \emph{one lightray at a time}. As a result, it will be better to work with the field strength components $\mathbf{\hat F}_{u\perp} = \del_u\mathbf{\hat A}_\perp$, which, unlike $\mathbf{\hat A}_\perp$ itself, don't depend on the gauge choice on \emph{neighboring} lightrays. With this in mind, let us use integration by parts w.r.t. $u$ to rewrite \eqref{eq:spinor_modes_final} in terms of field strengths:
\begin{align}
 \begin{split}
    c^+_{\text{out}}(\mu^\alpha,\tilde\mu_\alpha) &= -\frac{2i\langle\mu\boldsymbol{\sigma}\mu\rangle}{\langle\tilde\mu\mu\rangle^2}\cdot\int_{-\infty}^\infty du\,
    \mathbf{\hat F}_{u\perp}\!\left(u,0,-\frac{\langle\tilde\mu\boldsymbol{\sigma}\mu\rangle}{\langle\tilde\mu\mu\rangle}\right) e^{i\langle\tilde\mu\mu\rangle u/2} \ ; \\
    c^-_{\text{out}}(\mu^\alpha,\tilde\mu_\alpha) &= +\frac{2i\langle\tilde\mu\boldsymbol{\sigma}\tilde\mu\rangle}{\langle\tilde\mu\mu\rangle^2}\cdot\int_{-\infty}^\infty du\,
    \mathbf{\hat F}_{u\perp}\!\left(u,0,-\frac{\langle\tilde\mu\boldsymbol{\sigma}\mu\rangle}{\langle\tilde\mu\mu\rangle}\right) e^{i\langle\tilde\mu\mu\rangle u/2} \ .
\end{split} \label{eq:spinor_modes_final_F}    
\end{align}

The static-patch scattering problem now boils down to expressing the final mode coefficients $c^\pm_{\text{out}}(\mu,\tilde\mu)$ as functionals of the initial ones $c^\pm_{\text{in}}(\lambda,\tilde\lambda)$. 

\subsection{Working with fields vs. with Fock states over a vacuum} \label{sec:kinematics:flat:fields_vs_states}

In this section, we take a brief digression to discuss the relationship between our field-based formulation of the scattering problem, and the more conventional one based on Fock states. From our point of view, the more fundamental objects are the fields operators on the past and future horizons. From these, one may construct Fock states by taking two additional steps. The first step is to make a distinction between \emph{positive-frequency} and \emph{negative-frequency} field modes. In the context of our lightlike horizons, this can be done by choosing a lightlike time coordinate along the lightrays, and then Fourier-transforming with respect to that coordinate. Once defined, the positive-frequency and negative-frequency modes can be thought of as annihilation and creation operators. This designation then defines a \emph{vacuum state}, as the state that is annihilated by all the annihilation operators. Other states can now be formed by acting on the vacuum with creation operators, via the usual Fock procedure. Thus, the second step in going from field operators to states is to \emph{restrict attention to one of the two frequency signs}. Now, in principle, the two vacua as defined on the past and future horizons may or may not be the same. As we'll see below, in relevant cases for us, the two \emph{are} the same. When this is true, positive-frequency modes on the past horizon can only evolve into positive-frequency modes on the future horizon, and vice versa. This then completes the usual picture of Fock states evolving into Fock states. 

To recap, the entire vacuum/Fock-space structure follows from a choice of lightlike time coordinate on the horizons. There exist two particularly natural choices. The first is to use the coordinate $\tau = -\ln(-v)$ on the past horizon, and $\tau = \ln u$ on the future horizon. The range $\tau\in\bbR$ then covers the static patch, while translations in $\tau$ simply describe the static patch's time-translation symmetry, i.e. their generator is the static-patch Hamiltonian. Since these time translations are a spacetime symmetry, we are assured that the distinction into positive/negative frequencies, as well as the vacuum state, are consistent between the two horizons as anticipated above. With these structures, one can now define an S-matrix intrinsic to the static patch, which evolves Fock states into Fock states, and makes no reference to spacetime regions outside the patch. While this picture is conceptually simplest, it is technically inconvenient. The reason is the one mentioned in the Introduction: the static patch has no spatial translation symmetry, so one is forced to work instead with spherical harmonics. In \cite{Halpern:2015zia}, the static-patch S-matrix in this language was computed for a \emph{free} conformally-massless scalar, and the resulting expression is already quite non-trivial; extending it to include interactions seems prohibitively challenging. 

Because of this, our choice in this paper is to use the lightlike coordinates $v$ and $u$ themselves, instead of $\tau$. As discussed above, this entails the cost of extending the past/future horizons to causally cover the entire $dS_4$ spacetime, for the benefit of gaining spatial translation symmetry. The vacuum state defined by $v$ ($u$) on the past (future) horizon is the Bunch-Davies vacuum of global $dS_4$ (which will become the usual Minkowski vacuum upon conformal transformation to Minkowski spacetime). In particular, the modes \eqref{eq:vector_modes_initial} with $\omega>0$ define annihilation operators with respect to this vacuum, while the ones with $\omega<0$ define creation operators. While translations of $u$ and $v$ are \emph{not} spacetime symmetries, the associated choice of vacuum and distinction into positive/negative frequencies \emph{is} consistent between the two horizons. This is because the Bunch-Davies vacuum has a universal definition in terms of a Euclidean path integral.

Thus, in choosing $v$ and $u$ as the lightlike coordinates that get Fourier-transformed, we've effectively taken the first step on the way from field modes to Fock states: we made a choice of vacuum, which is conformal to the usual Minkowski one. As a result, our scattering formulas will look a lot like ordinary S-matrix amplitudes with respect to the usual Minkowski vacuum. However, this vacuum is a pure state of \emph{global} $dS_4$, rather than of the static patch, and the associated positive/negative frequency modes are only defined on the extended horizons, rather than on their observable ``halves''. As a result, we choose to \emph{not} take the second step, of throwing away the field modes with the ``wrong'' frequency sign. Instead, we consider arbitrary field modes on the observed half of e.g. the past horizon, and then extend them arbitrarily into the unobservable half.

We conclude this subsection with a final comment. The word ``scattering'' in Minkowski space is often associated with the notion of fields becoming free as one approaches $\scri^\pm$. This is often presented as a prerequisite for the Fock-space construction of particle states. On the other hand, near a de Sitter horizon, the fields are no more free than anywhere else in the spacetime. There is in fact no conflict here. For fields on a null hypersurface, one can always perform the Fock-space construction based on positive/negative frequencies with respect to a lightlike coordinate. Interactions do not affect this picture. From this point of view, the asymptotic boundary $\scri^\pm$ of Minkowski space is just another null hypersurface. However, since it's only defined asymptotically, one may worry whether the fields will have well-defined values on it. It is \emph{this} requirement that translates into the need for Minkowski fields to be ``sufficiently free'' near $\scri^\pm$. For horizons in the bulk of $dS_4$, the issue does not arise. 

\subsection{Flat coordinates adapted to past horizon} \label{sec:kinematics:flat}

\begin{figure}%
	\centering%
	\includegraphics[scale=1]{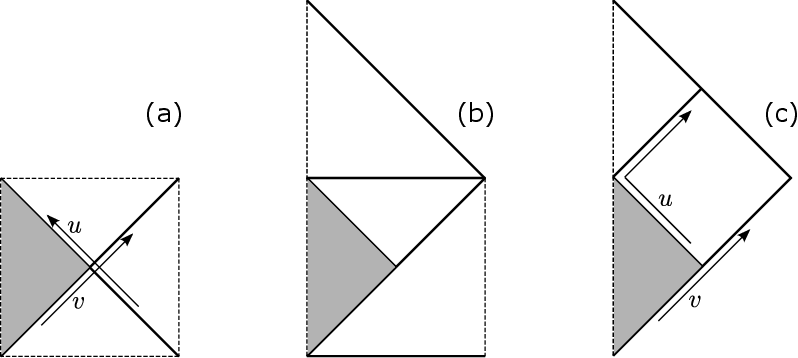} \\
	\caption{Penrose diagrams of the $dS_4$ static-patch problem and its Minkowski counterpart. (a) The static patch (in gray) inside $dS_4$; its past and future lightlike boundaries have been ``doubled'' into geodesically complete cosmological horizons. (b) The $dS_4$ Poincare patch associated with the past horizon, and its extension into a full Minkowski space through the future boundary of $dS_4$. The static patch, still in gray, is a subregion of both $dS_4$ and Minkowski. (c) The Minkowski picture, with $dS_4$ removed. The static patch is now the causal past of the origin, the past horizon is past lightlike infinity, and the future horizon is the origin's lightcone. This lightcone is again a geodesically complete ``doubling'' of the static patch's future boundary, but now the completion is into the future.}
	\label{fig:Penrose} 
\end{figure}%
Having defined the static-patch problem in section \ref{sec:kinematics:static}, we will now set up a flat conformal frame in which it can be solved more easily. This frame will be based on the Poincare coordinates associated with the \emph{past} horizon, breaking the symmetry between the two horizons which we've been careful to maintain so far. These Poincare coordinates $x^\mu = (t,\mathbf{x})$ are related to the embedding-space coordinates $r^I = (u,v,\mathbf{r})$ of section \ref{sec:kinematics:static} as:
\begin{align}
 (u,v,\mathbf{r}) = -\frac{1}{t}\left(1, \frac{\mathbf{x}^2 - t^2}{2}, \mathbf{x}\right) \ . \label{eq:Poincare}
\end{align}
The $x^\mu$ coordinates define a flat metric $\eta_{\mu\nu}dx^\mu dx^\nu \equiv -dt^2 + \mathbf{dx}^2$, which is conformally related to the $dS_4$ metric \eqref{eq:dS_metric} via:
\begin{align}
 -2dudv + \mathbf{dr}^2 = \frac{1}{t^2}\,\eta_{\mu\nu}dx^\mu dx^\nu \ .
\end{align}
The YM gauge potential is conformally invariant. We may therefore disregard the conformal factor of $t^2$, and work with the Minkowski metric $\eta_{\mu\nu}$. The potential's components $A_\mu = (A_t,\mathbf{A})$ in the $x^\mu$ basis are related to those in the embedding-space basis via $A_\mu = (\del r^I/\del x^\mu)\hat A_I$. Imposing the constraint $r^I \hat A_I = 0$, the relation between the components becomes:
\begin{align}
  \hat A_u = \frac{1}{2}(t^2 + \mathbf{x}^2)A_t + t(\mathbf{x\cdot A}) \ ; \quad \hat A_v = A_t \ ; \quad \mathbf{\hat A} = -t\mathbf{A} - \mathbf{x}A_t \ . \label{eq:A_transformation}
\end{align}
The field strength in the intrinsic $x^\mu$ coordinates is derived from the potential as:
\begin{align}
 F_{\mu\nu} = 2(\del_{[\mu}A_{\nu]} + A_{[\mu} A_{\nu]}) \ . \label{eq:F}
\end{align}
The coordinate range $t<0$ spans the expanding Poincare patch $u>0$ of $dS_4$, i.e. the half of $dS_4$ that lies to the future of the past horizon. The past horizon itself is given by the limit $t\rightarrow-\infty,|\mathbf{x}|\rightarrow\infty$ with $t+|\mathbf{x}|$ finite. In particular, the coordinates $(v,\mathbf{r})$ on the initial horizon are given in this limit by:
\begin{align}
 v = t+|\mathbf{x}| \ ; \quad \mathbf{r} = \frac{\mathbf{x}}{|\mathbf{x}|} \quad \text{(with }t\rightarrow-\infty\text{, }|\mathbf{x}|\rightarrow\infty\text{)} \ . \label{eq:past_horizon}
\end{align}
Thus, in the coordinates $x^\mu$, the past horizon becomes past lightlike infinity $\scri^-$. As discussed in section \ref{sec:kinematics:static}, the actual past boundary of the static patch is restricted to $v<0$, but we extend the initial data to the entire range $v\in\bbR$ arbitrarily. 

The \emph{future} horizon of the $dS_4$ static patch becomes the origin's lightcone $\eta_{\mu\nu}x^\mu x^\nu = 0$ in the $x^\mu$ coordinates. The horizon coordinates $(u,\mathbf{r})$ are then given in terms of $x^\mu$ as:
\begin{align}
 u = \pm\frac{1}{|\mathbf{x}|} \ ; \quad \mathbf{r} = \pm\frac{\mathbf{x}}{|\mathbf{x}|} \quad \text{(with }t = \mp|\mathbf{x}|\text{, respectively)} \ . \label{eq:future_horizon}
\end{align}
In particular, the observable half-horizon $u>0$ is described by the past lightcone $t=-|\mathbf{x}|$. See the Penrose diagrams in Figure \ref{fig:Penrose}(b,c).

Note that as we switch conformal frames between $dS_4$ and the Minkowski space $x^\mu$, the spacetime's global structure changes. As the Poincare time $t$ increases through the range $t<0$, it reaches the future conformal boundary of $dS_4$ at $t=0^-$. We then encounter a discontinuity: $t=0^+$ is at the \emph{past} conformal boundary of $dS_4$, and the range $t>0$ spans a \emph{contracting} Poincare patch that lies to the \emph{past} of the past horizon. In contrast, from the point of view of the flat metric $\eta_{\mu\nu}$, $t=0$ is just a regular time slice, and the flat Minkowski space continues right through it. On the other hand, the flat metric treats the past horizon as $\scri^-$, and doesn't see the complementary Poincare patch to its past. In particular, both conformal frames agree that the observable half $u>0$ of the future horizon is geodesically incomplete, and that completing it involves extending $u$ or $t=-1/u$ to the entire real line. However, the two frames disagree on the \emph{direction} of this extension. In $dS_4$, the horizon wants to continue smoothly into the past, from $u>0$ to $u<0$, which for the $t$ coordinate looks like a discontinuous jump through $t=\pm\infty$. In Minkowski, the picture is reversed: the horizon (or, rather, lightcone) wants to continue smoothly into the \emph{future}, from $t<0$ to $t>0$, which is discontinuous for the $u$ coordinate. Fundamentally, it doesn't matter which of the pictures we adopt, since they only disagree \emph{outside} the observable static patch (though note that the term ``observable'' here still refers to the $dS_4$ metric). As a matter of \emph{convenience}, we will adopt the Minkowski frame, and with it the extension of the future horizon into the future through $t=0$, rather than into the past through $u=0$. We can continue using the formulas \eqref{eq:spinor_modes_final}-\eqref{eq:spinor_modes_final_F} for the future horizon modes, but with the understanding that the $u<0$ range refers to the future $t>0$ half of Minkowski space, rather than to the contracting Poincare patch of $dS_4$. 

We now turn to introduce spinor notation for the Minkowski space $x^\mu$. This simply extends the 3d spinor notation from section \ref{sec:kinematics:static}. In particular, we introduce a distinction between left-handed (undotted) and right-handed (dotted) spinor indices. Index raising and lowering are defined as before:
\begin{align}
 \psi_\alpha = \epsilon_{\alpha\beta}\psi^\beta \ ; \quad \psi^\beta = \psi_\alpha\epsilon^{\alpha\beta} \ ; \quad \tilde\psi_{\dot\alpha} = \epsilon_{\dot\alpha\dot\beta}\tilde\psi^{\dot\beta} \ ; \quad \tilde\psi^{\dot\beta} = \tilde\psi_{\dot\alpha}\epsilon^{\dot\alpha\dot\beta} \ ,
\end{align}
and we define shorthands for inner products as:
\begin{align}
 \psi_\alpha\chi^\alpha \equiv \langle\psi\chi\rangle \ ; \quad \tilde\psi_{\dot\alpha}\tilde\chi^{\dot\alpha} \equiv [\tilde\psi\tilde\chi] \ ; \quad \psi_\alpha V^\alpha{}_{\dot\alpha}\tilde\chi^{\dot\alpha} \equiv \langle\psi V\tilde\chi] \ .
\end{align}
Spinor complex conjugation is now defined by $\bar\psi^{\dot\alpha} = (\psi^\alpha)^*$. The 3d Pauli matrices become $\boldsymbol{\sigma}^{\alpha\dot\alpha}$, and are incorporated with the identity matrix $\sigma_t^{\alpha\dot\alpha}$ into the 4d Pauli matrices $\sigma_\mu^{\alpha\dot\alpha}$, which satisfy:
\begin{align}
 \sigma^\mu_{\alpha\dot\alpha}\sigma_\nu^{\alpha\dot\alpha} = -2\delta^\mu_\nu \ ; \quad \sigma_\mu^{\alpha\dot\alpha}\sigma^\mu_{\beta\dot\beta} = -2\delta^\alpha_\beta \delta^{\dot\alpha}_{\dot\beta} \ ; \quad
 \sigma_{(\mu}^{\alpha\dot\alpha} \sigma_{\nu)\beta\dot\alpha} = -\eta_{\mu\nu}\delta^\alpha_\beta \ ; \quad \sigma_{(\mu}^{\alpha\dot\alpha} \sigma_{\nu)\alpha\dot\beta} = -\eta_{\mu\nu}\delta^{\dot\alpha}_{\dot\beta} \ .
\end{align}
We use $\sigma_\mu^{\alpha\dot\alpha}$ to translate between vector and spinor indices, via:
\begin{align}
 V^{\alpha\dot\alpha} = V^\mu\sigma_\mu^{\alpha\dot\alpha} \ ; \quad V^\mu = -\frac{1}{2}V^{\alpha\dot\alpha}\sigma^\mu_{\alpha\dot\alpha} \ .
\end{align}
The YM gauge potential \eqref{eq:A_transformation} can now be written as $A_{\alpha\dot\alpha}$. Its field strength \eqref{eq:F} decomposes as:
\begin{align}
 F_{\alpha\dot\alpha\beta\dot\beta} = \epsilon_{\alpha\beta}F_{\dot\alpha\dot\beta} + \epsilon_{\dot\alpha\dot\beta}F_{\alpha\beta} \ .
\end{align}
$F_{\dot\alpha\dot\beta}$ and $F_{\alpha\beta}$ encode the self-dual (right-handed) and anti-self-dual (left-handed) parts of $F_{\mu\nu}$, respectively. In terms of $A_{\alpha\dot\alpha}$, they read:
\begin{align}
 F_{\alpha\beta} &= \del_{(\alpha}{}^{\dot\alpha}A_{\beta)\dot\alpha} + A_{(\alpha}{}^{\dot\alpha}A_{\beta)\dot\alpha} \ ; \label{eq:F_left} \\
 F_{\dot\alpha\dot\beta} &= -\del_{\alpha(\dot\alpha}A^\alpha{}_{\dot\beta)} - A_{\alpha(\dot\alpha}A^\alpha{}_{\dot\beta)} \ , \label{eq:F_right}
\end{align}
where $\del_{\alpha\dot\alpha}\equiv \sigma^\mu_{\alpha\dot\alpha}\del_\mu$.

\subsection{Lightlike plane waves and initial horizon data} \label{sec:kinematics:linear}

A lightlike momentum $k^\mu=(\omega,\mathbf{k})$, either future-pointing or past-pointing, can be written in spinor notation as:
\begin{align}
 k_\mu = \frac{1}{2}\langle\lambda\sigma_\mu\tilde\lambda] \ ; \quad k_{\alpha\dot\alpha} = \lambda_\alpha\tilde\lambda_{\dot\alpha} \ ; \quad \tilde\lambda_{\dot\alpha} = \sign(\omega)\bar\lambda_{\dot\alpha} \ . \label{eq:spinor_helicity_4d}
\end{align}
This coincides with our previous $SO(3)$-spinor expression \eqref{eq:spinor_helicity_3d}, given the componentwise equality of the 4d complex conjugate $\bar\lambda^{\dot\alpha}$ and the 3d one $\bar\lambda_\alpha$. The momentum \eqref{eq:spinor_helicity_4d} is again invariant under the phase rotations:
\begin{align}
 \lambda^\alpha \rightarrow e^{i\phi}\lambda^\alpha \ ; \quad \tilde\lambda^{\dot\alpha} \rightarrow e^{-i\phi}\tilde\lambda^{\dot\alpha} \ . \label{eq:phase_rotation_4d}
\end{align}
A lightlike wave with the momentum \eqref{eq:spinor_helicity_4d} takes the form:
\begin{align}
 e^{ik\cdot x} \equiv e^{ik_\mu x^\mu} = e^{i\langle\lambda x\tilde\lambda]/2} \ ; \quad \del_{\alpha\dot\alpha}e^{ik\cdot x} = ik_{\alpha\dot\alpha}e^{ik\cdot x} = i\lambda_\alpha\tilde\lambda_{\dot\alpha}e^{ik\cdot x} \ . \label{eq:plane_wave}
\end{align}
Adding appropriate polarization factors, we can construct purely right-handed or left-handed plane-wave solutions to the Maxwell equations:
\begin{align}
 A_{\alpha\dot\alpha} = -i\frac{q_\alpha\tilde\lambda_{\dot\alpha}}{\langle q\lambda\rangle}\,e^{ik\cdot x} \ &\Longrightarrow \  
   F_{\dot\alpha\dot\beta} = \tilde\lambda_{\dot\alpha}\tilde\lambda_{\dot\beta}e^{ik\cdot x} \ ; \quad F_{\alpha\beta} = 0 \ ; \\
 A_{\alpha\dot\alpha} = -i\frac{\lambda_\alpha \tilde q_{\dot\alpha}}{[\tilde q\tilde\lambda]}\,e^{ik\cdot x} \ &\Longrightarrow \ 
   F_{\alpha\beta} = \lambda_\alpha\lambda_\beta e^{ik\cdot x} \ ; \quad F_{\dot\alpha\dot\beta} = 0 \ , \label{eq:A_F_plane_wave}
\end{align}
where $q^\alpha$ and $\tilde q^{\dot\alpha}$ are arbitrary spinors encoding the gauge freedom (in particular, the field strength doesn't depend on them). The general \emph{linearized} solution to the YM equations is obtained by integrating over such plane waves with gauge-algebra-valued coefficients $c^\pm_{\text{in}}(\lambda,\tilde\lambda)$:
\begin{align}
 A_{\alpha\dot\alpha}^{\text{lin}}(x^\mu) &= -\frac{i}{2\pi^2}\int_{k^2 = 0} \frac{d^3\mathbf{k}}{2\omega}
  \left(\frac{q_\alpha\tilde\lambda_{\dot\alpha}}{\langle q\lambda\rangle}\,c^+_{\text{in}}(\lambda^{\beta},\tilde\lambda^{\dot\beta}) 
    + \frac{\lambda_\alpha \tilde q_{\dot\alpha}}{[\tilde q\tilde\lambda]}\,c^-_{\text{in}}(\lambda^{\beta},\tilde\lambda^{\dot\beta}) \right) e^{ik\cdot x} \ ; \label{eq:linearized_A} \\
 F_{\dot\alpha\dot\beta}^{\text{lin}}(x^\mu) &= \frac{1}{2\pi^2}\int_{k^2 = 0} \frac{d^3\mathbf{k}}{2\omega}\,
  \tilde\lambda_{\dot\alpha}\tilde\lambda_{\dot\beta}\,c^+_{\text{in}}(\lambda^{\beta},\tilde\lambda^{\dot\beta})\,e^{ik\cdot x} \ ; \label{eq:linearized_F_right} \\
 F_{\alpha\beta}^{\text{lin}}(x^\mu) &= \frac{1}{2\pi^2}\int_{k^2 = 0} \frac{d^3\mathbf{k}}{2\omega}\,
   \lambda_\alpha\lambda_\beta\,c^-_{\text{in}}(\lambda^{\beta},\tilde\lambda^{\dot\beta})\,e^{ik\cdot x} \ , \label{eq:linearized_F_left} 
\end{align}
where the integration range is understood to include both positive-frequency and negative-frequency modes:
\begin{align}
 \int_{k^2 = 0} \equiv \int_{\omega=|\mathbf{k}|} + \int_{\omega=-|\mathbf{k}|} \ .
\end{align}
The spinors $(\lambda^\alpha,\tilde\lambda^{\dot\alpha})$ in the integrand of \eqref{eq:linearized_A} are the square root of $k^\mu$, as in \eqref{eq:spinor_helicity_4d}, with the phase freedom \eqref{eq:phase_rotation_4d} fixed arbitrarily. For the result to not depend on the choice of phase, the mode coefficients $c^\pm_{\text{in}}$ must transform under the phase rotations \eqref{eq:phase_rotation_4d} with the appropriate weights $\pm 2$:
\begin{align}
 c^\pm_{\text{in}}(e^{i\phi}\lambda, e^{-i\phi}\tilde\lambda) = e^{\pm 2i\phi} c^\pm_{\text{in}}(\lambda, \tilde\lambda) \ .
\end{align}
Note that we used the same notation $c^\pm_{\text{in}}$ for the plane-wave coefficients in \eqref{eq:linearized_A} and for the mode coefficients \eqref{eq:spinor_modes_initial} on the past horizon. Let us show that they are in fact equal, justifying our identification of \eqref{eq:k} as a 4-momentum. To do this, we evaluate the potential \eqref{eq:linearized_A} in the null-infinity limit \eqref{eq:past_horizon} that describes the past horizon in the $x^\mu$ coordinates. This is a standard calculation, in which we decompose the $d^3\mathbf{k}$ integral into integrals over its magnitude $\mathbf{k}$ and its direction $\mathbf{k}/|\mathbf{k}|$. In the null-infinity limit, the integral over directions can be found by the stationary-phase method, with the two stationary points $\mathbf{k} = \pm\omega\mathbf{r}$. Of these, only the point $\mathbf{k} = -\omega\mathbf{r}$ survives; the other leads to a rapidly oscillating phase in the integral over the magnitude $|\mathbf{k}|$. All in all, we get:
\begin{align}
  A_\mu = -\frac{1}{4\pi t}\int_{-\infty}^\infty d\omega \left( \frac{\langle q\sigma_\mu\tilde\lambda]}{\langle q\lambda\rangle}\,c^+_{\text{in}}(\lambda^{\beta},\tilde\lambda^{\dot\beta}) 
    - \frac{\langle\lambda\sigma_\mu\tilde q]}{[\tilde\lambda\tilde q]}\,c^-_{\text{in}}(\lambda^{\beta},\tilde\lambda^{\dot\beta}) \right) e^{-i\omega v} \ , \label{eq:A_past_horizon}
\end{align}
where the spinors $(\lambda,\tilde\lambda)$ are related to $(\omega,\mathbf{r})$ as in \eqref{eq:k}-\eqref{eq:spinor_helicity_3d}, and $t\approx-|\mathbf{x}|$ goes to $-\infty$. Due to the $t$ in the denominator, the components $A_\mu$ on the past horizon all vanish. Transforming into the embedding-space basis via \eqref{eq:A_transformation}, we conclude that we are automatically in the gauge $\hat A_v = 0$ in which the initial data \eqref{eq:spinor_modes_initial} is defined. As for the spatial components $\mathbf{\hat A}_\perp$ along the 2-sphere in the embedding-space basis, they are given by $-t$ times the corresponding components of \eqref{eq:A_past_horizon}, which is finite. Furthermore, these transverse components don't depend on the choice of gauge spinors $(q,\tilde q)$, which can only shift $A_\mu$ longitudinally, along $\langle\lambda\sigma_\mu\tilde\lambda]$. To make contact with the $SO(3)$ formalism of section \ref{sec:kinematics:static}, it's convenient to choose:
\begin{align}
 q^\alpha = \sigma_t^{\alpha\dot\alpha}\tilde\lambda_{\dot\alpha} \ ; \quad \tilde q^{\dot\alpha} = \sigma_t^{\alpha\dot\alpha}\lambda_\alpha \ ,
\end{align}
which is equivalent to fixing $A_t=0$ everywhere. We can now descend to 3d spinor notation, leaving only undotted spinor indices, treating $\sigma_t^{\alpha\dot\alpha}$ as the identity matrix, and identifying $\tilde\lambda^{\dot\alpha}$ with $\tilde\lambda_\alpha$. The potential's spatial components on the past horizon then read:
\begin{align}
 \mathbf{\hat A}(0,v,\mathbf{r}) = \frac{1}{4\pi}\int_{-\infty}^\infty d\omega \left( \frac{\langle\tilde\lambda\boldsymbol{\sigma}\tilde\lambda\rangle}{\langle\tilde\lambda\lambda\rangle}\,c^+_{\text{in}}(\lambda^\beta,\tilde\lambda_\beta) 
  - \frac{\langle\lambda\boldsymbol{\sigma}\lambda\rangle}{\langle\tilde\lambda\lambda\rangle}\,c^-_{\text{in}}(\lambda^\beta,\tilde\lambda_\beta) \right) e^{-i\omega v} \ ,
\end{align}
where we recognize the null polarization vectors from \eqref{eq:m}. Contracting with these vectors and Fourier-transforming with respect to $v$, we recover the initial-data expressions \eqref{eq:spinor_modes_initial}. 

\subsection{Non-linear corrections and final horizon data} \label{sec:kinematics:non_linear}

Ultimately, we are interested in the final data $c^\pm_{\text{out}}(\mu,\tilde\mu)$ as a functional of the initial data $c^\pm_{\text{in}}(\lambda,\tilde\lambda)$. The Taylor coefficients of this functional define the static-patch ``scattering amplitudes'', which we'll denote as $\calS(1^{h_1},\dots,n^{h_n};\mu,\tilde\mu,h)$. Here, $n$ is the number of ingoing $c^\pm_{\text{in}}$ factors, and each argument $i^{h_i}$ is a shorthand for a pair of spinor-helicity variables $(\lambda_i,\tilde\lambda_i)$, along with a helicity sign $h_i=\pm$:
\begin{align}
 i^{h_i} \equiv \{\lambda_i^\alpha,\tilde\lambda_i^{\dot\alpha},h_i \} \ .
\end{align}
Similarly, $(\mu,\tilde\mu)$ are the spinor-helicity variables for the outgoing $c^\pm_{\text{out}}$ mode, and $h=\pm$ is its helicity sign. For brevity, we will sometimes omit the dependence on $(\mu,\tilde\mu)$. Note that the order of the ingoing legs $(1,\dots,n)$ is important, because the $c^\pm_{\text{in}}$ initial data are gauge-algebra-valued, and thus do not commute. As mentioned above, we follow here the ``color ordering'' convention, which is to treat each ordering of $c^\pm_{\text{in}}$ as a distinct term. With this convention, the group's structure constants never enter the calculation, and the color-ordered ``amplitudes'' $\calS$ (which themselves are gauge singlets) do not depend on the gauge group. Note that since the modes $c^\pm_{\text{in}}$ contain both positive and negative energies, they may also not commute as quantum operators; however, we will work at tree level, where this issue doesn't arise. Our expression for $c^\pm_{\text{out}}$ in terms of the ``amplitudes'' $\calS$ is given below, in eq. \eqref{eq:kinematics}. To motivate the prefactors there, we must first prepare some groundwork.

At tree level, the final data $c^\pm_{\text{out}}$ can be read off from a classical field solution $A_{\alpha\dot\alpha}(x)$, determined by the initial data $c^\pm_{\text{in}}$. We will therefore need the non-linear corrections to the linearized potential \eqref{eq:linearized_A}. From now on, we mostly specialize to a gauge in which the spinors $(q,\tilde q)$ in \eqref{eq:linearized_A} are constant, i.e. do not depend on $(\lambda,\tilde\lambda)$. This amounts to the gauge condition:
\begin{align}
 \langle q A \tilde q] = 0 \ , \label{eq:q_gauge}
\end{align} 
i.e. a lightcone gauge with respect to the constant null vector $q^\alpha\tilde q^{\dot\alpha}$. We now apply the condition \eqref{eq:q_gauge} to the entire non-linear field, thus gauge-fixing the non-linear corrections to \eqref{eq:linearized_A}. Making the dependence on $(q,\tilde q)$ explicit, we Taylor-expand the potential $A_{\alpha\dot\alpha}$ in powers of the initial data $c^\pm_{\text{in}}$ as:
\begin{align}
 A_{\alpha\dot\alpha}(x;q,\tilde q) = \sum_{n=1}^\infty \int_{1\dots n} e^{iK_{1\dots n}\cdot x}\sum_{h_1,\dots,h_n} a_{\alpha\dot\alpha}(1^{h_1},\dots,n^{h_n};q,\tilde q)
     \,c^{h_1}_{\text{in}}(\lambda_1,\tilde\lambda_1)\dots c^{h_n}_{\text{in}}(\lambda_n,\tilde\lambda_n) \ . \label{eq:A_expansion}
\end{align}
Here, we sum over all choices of the ingoing helicity signs $(h_1,\dots,h_n)$. We use $\int_{1\dots n}$ as a shorthand for an $n$-fold integral over on-shell ingoing momenta as in \eqref{eq:linearized_A}-\eqref{eq:linearized_F_left}:
\begin{align}
 \int_{1\dots n} \equiv \left(\frac{1}{2\pi^2}\int_{k_1^2 = 0} \frac{d^3\mathbf{k}_1}{2\omega_1}\right)\ldots\left(\frac{1}{2\pi^2}\int_{k_n^2 = 0} \frac{d^3\mathbf{k}_n}{2\omega_n}\right) \ .
\end{align}
$K_{1\dots n}$ denotes the sum of these momenta:
\begin{align}
 K^{\alpha\dot\alpha}_{1\dots n} \equiv k_1^{\alpha\dot\alpha} + \ldots + k_n^{\alpha\dot\alpha} = \lambda_1^\alpha\tilde\lambda_1^{\dot\alpha} + \ldots + \lambda_n^\alpha\tilde\lambda_n^{\dot\alpha} \ , \label{eq:K_sum}
\end{align}
and we will similarly denote partial sums of consecutive momenta by $K^\mu_{i\dots j}$.

At $n=1$, the coefficients $a_{\alpha\dot\alpha}$ in \eqref{eq:A_expansion} can be read off from the linearized potential \eqref{eq:linearized_A} as:
\begin{align}
 a_{\alpha\dot\alpha}(+;q,\tilde q) = -i\frac{q_\alpha\tilde\lambda_{\dot\alpha}}{\langle q\lambda\rangle} \ ; \quad a_{\alpha\dot\alpha}(-;q,\tilde q) = -i\frac{\lambda_\alpha \tilde q_{\dot\alpha}}{[\tilde q\tilde\lambda]} \ . \label{eq:linearized_a}
\end{align}
The coefficients with $n\geq 2$ describe the non-linear corrections to the bulk field, which can be found by computing Feynman diagrams with $n+1$ external legs, of which $n$ are on-shell and 1 is off-shell. When the momentum \eqref{eq:K_sum} of the off-shell, ``outgoing'' leg goes on-shell, the coefficients $a_{\alpha\dot\alpha}$ acquire poles, whose residues are the Minkowski scattering amplitudes. We will discuss these in section \ref{sec:poles:S_matrix}. However, generally, we will need not only these residues, but the non-linear field itself, at general, off-shell momenta $K^\mu$.

As usual, to be uniquely defined, the non-linear corrections require boundary conditions, which amount to an $i\varepsilon$ prescription in the propagators. We fix these by demanding that the initial data $c^\pm_{\text{in}}(\lambda,\tilde\lambda)$ continues to describe the field at past null infinity, i.e. at the past horizon, in the sense of \eqref{eq:spinor_modes_initial}. This dictates that we should use retarded propagators, which can be encoded by adding an infinitesimal future-pointing imaginary part to each ingoing 4-momentum $k^\mu$. We will keep this understanding implicit, and omit $i\varepsilon$'s below.

In complete analogy with \eqref{eq:A_expansion}, we define expansions of the right-handed and left-handed field strengths:
\begin{align}
 F_{\dot\alpha\dot\beta}(x;q,\tilde q) &= \sum_{n=1}^\infty \int_{1\dots n} e^{iK_{1\dots n}\cdot x} \sum_{h_1,\dots,h_n} f_{\dot\alpha\dot\beta}(1^{h_1},\dots,n^{h_n};q,\tilde q)\,
    c^{h_1}_{\text{in}}(\lambda_1,\tilde\lambda_1)\dots c^{h_n}_{\text{in}}(\lambda_n,\tilde\lambda_n) \ ; \label{eq:F_right_expansion} \\
 F_{\alpha\beta}(x;q,\tilde q) &= \sum_{n=1}^\infty \int_{1\dots n} e^{iK_{1\dots n}\cdot x} \sum_{h_1,\dots,h_n} f_{\alpha\beta}(1^{h_1},\dots,n^{h_n};q,\tilde q)\,
    c^{h_1}_{\text{in}}(\lambda_1,\tilde\lambda_1)\dots c^{h_n}_{\text{in}}(\lambda_n,\tilde\lambda_n) \ . \label{eq:F_left_expansion}
\end{align}
At $n=1$, the field strength's coefficients can be read off from \eqref{eq:A_F_plane_wave},\eqref{eq:linearized_a} as:
\begin{align}
 \begin{split}
   f_{\dot\alpha\dot\beta}(+;q,\tilde q) &= \tilde\lambda_{\dot\alpha}\tilde\lambda_{\dot\beta} \ ; \quad f_{\alpha\beta}(-;q,\tilde q) = \lambda_\alpha\lambda_\beta \ ; \\
   f_{\dot\alpha\dot\beta}(-;q,\tilde q) &= f_{\alpha\beta}(+;q,\tilde q) = 0 \ . 
 \end{split} \label{eq:f_linear} 
\end{align}
While these coefficients of the linearized field strength don't depend on the gauge spinors $(q,\tilde q)$, this will not be the case for the non-linear corrections.

Let us now understand how the final data $c^\pm_{\text{out}}$ on the future horizon can be read off from the non-linear fields \eqref{eq:A_expansion} or \eqref{eq:F_right_expansion}-\eqref{eq:F_left_expansion}. Unlike the initial data, which is unaffected by the non-linear corrections, the final data will receive contributions from both the linear and non-linear terms in \eqref{eq:A_expansion}. The linear contribution can be worked out using a spinor Fourier transform \cite{David:2019mos}, as was done explicitly for a scalar field in \cite{Albrychiewicz:2020ruh}. Here, we'll present an alternative derivation, which works equally well for the off-shell momenta of the non-linear corrections. Consider the final data $c^\pm_{\text{out}}(\mu,\tilde\mu)$ on the final horizon, evaluated at some value of the spinor-helicity variables $(\mu,\tilde\mu)$. In embedding-space coordinates, this is given by a Fourier transform \eqref{eq:spinor_modes_final}-\eqref{eq:spinor_modes_final_F} with respect to the null time $u$, along the lightray $\mathbf{r} = -\langle\tilde\mu\boldsymbol{\sigma}\mu\rangle/\langle\tilde\mu\mu\rangle$. In our flat frame, the future horizon corresponds to the origin's lightcone, as in eq. \eqref{eq:future_horizon}. Thus, in Minkowski coordinates, the lightray along which the Fourier transform \eqref{eq:spinor_modes_final}-\eqref{eq:spinor_modes_final_F} is taken reads:
\begin{align}
 x^\mu = -\frac{1}{u}\left(1,\frac{\langle\tilde\mu\boldsymbol{\sigma}\mu\rangle}{\langle\tilde\mu\mu\rangle} \right) = -\frac{\langle\mu\sigma^\mu\tilde\mu]}{\langle\tilde\mu\mu\rangle u} \ . \label{eq:mu_lightray}
\end{align}
As for our gauge choice $\hat A_u = 0$ on the future horizon, it becomes simply $\langle\mu A\tilde\mu] = 0$. The simplest way to impose this gauge condition on the lightray \eqref{eq:mu_lightray} is to impose it \emph{everywhere}, i.e. to adopt the final-horizon spinors $(\mu,\tilde\mu)$ as our gauge spinors $(q,\tilde q)$, defining the lightcone gauge \eqref{eq:q_gauge}. Thus, to evaluate the final data on each separate lightray of the future horizon, we will calculate the bulk field $A_{\alpha\dot\alpha}(x)$ in a separate lightcone gauge. Note that this doesn't affect our encoding $c^\pm_{\text{in}}(\lambda,\tilde\lambda)$ of the initial data, since the latter doesn't depend on $(q,\tilde q)$. However, we can no longer use the formula \eqref{eq:spinor_modes_final} for the final modes, because the potential $A_\mu$ depends on the gauge choice not only on the \emph{chosen} lightray, but also on the \emph{neighboring} ones. Instead, we must use the formula \eqref{eq:spinor_modes_final_F}, which uses the field strength.

Let us now see exactly how $c^\pm_{\text{out}}(\mu,\tilde\mu)$ can be read off from the non-linear bulk potential in the appropriate gauge, i.e. from $A_{\alpha\dot\alpha}(x;\mu,\tilde\mu)$. For the moment, we can abstract away from the expansion \eqref{eq:A_expansion}, and simply consider $A_{\alpha\dot\alpha}$ and its field strength in momentum space, i.e. decomposed into general plane waves:
\begin{align}
 A_{\alpha\dot\alpha}(x;\mu,\tilde\mu) &\equiv \int d^4K\,\calA_{\alpha\dot\alpha}(K;\mu,\tilde\mu)\,e^{iK\cdot x} \ ; \label{eq:A_Fourier} \\
 F_{\alpha\beta}(x;\mu,\tilde\mu) &\equiv \int d^4K\,\calF_{\alpha\beta}(K;\mu,\tilde\mu)\,e^{iK\cdot x} \ ; \label{eq:F_left_Fourier} \\ 
 F_{\dot\alpha\dot\beta}(x;\mu,\tilde\mu) &\equiv \int d^4K\,\calF_{\dot\alpha\dot\beta}(K;\mu,\tilde\mu)\,e^{iK\cdot x} \ . \label{eq:F_right_Fourier}
\end{align}
Let's now plug this field into our definition \eqref{eq:spinor_modes_final_F} of the final data. Eq. \eqref{eq:spinor_modes_final_F} refers to the field strength in the embedding-space basis. The relevant components are related to those in the Minkowski basis via:
\begin{align}
 \langle\mu\boldsymbol{\sigma}\mu\rangle\cdot\mathbf{\hat F}_{u\perp} = -\frac{1}{u^3}\mu^\alpha\mu^\beta F_{\alpha\beta} \ ; \quad
 \langle\tilde\mu\boldsymbol{\sigma}\tilde\mu\rangle\cdot\mathbf{\hat F}_{u\perp} = \frac{1}{u^3}\tilde\mu^{\dot\alpha}\tilde\mu^{\dot\beta} F_{\dot\alpha\dot\beta} \ . \label{eq:F_components}
\end{align}
These helicity components can also be expressed in terms of the (Minkowski, lightcone gauge) potential $A_{\alpha\dot\alpha}$. Indeed, in the gauge $\langle\mu A\tilde\mu] = 0$, we have $\mu^\alpha A_{\alpha\dot\alpha}\sim\tilde\mu_{\dot\alpha}$ and $\tilde\mu^{\dot\alpha}A_{\alpha\dot\alpha}\sim\mu_{\alpha}$, which implies the vanishing of $\mu^\alpha\mu^\beta A_{(\alpha}{}^{\dot\alpha}A_{\beta)\dot\alpha}$ and $\tilde\mu^{\dot\alpha}\tilde\mu^{\dot\beta} A_{\alpha(\dot\alpha}A^\alpha{}_{\dot\beta)}$. As a result, the field-strength components $\mu^\alpha\mu^\beta F_{\alpha\beta}$ and $\tilde\mu^{\dot\alpha}\tilde\mu^{\dot\beta} F_{\dot\alpha\dot\beta}$ depend on the potential \emph{linearly}:
\begin{align}
 \begin{split}
   \mu^\alpha\mu^\beta F_{\alpha\beta}(x;\mu,\tilde\mu) &= \mu^\alpha\mu^\beta\del_\alpha{}^{\dot\alpha}A_{\beta\dot\alpha}(x;\mu,\tilde\mu) \ ; \\
   \tilde\mu^{\dot\alpha}\tilde\mu^{\dot\beta} F_{\dot\alpha\dot\beta}(x;\mu,\tilde\mu) &= \tilde\mu^{\dot\alpha}\tilde\mu^{\dot\beta} \del^\alpha{}_{\dot\alpha}A_{\alpha\dot\beta}(x;\mu,\tilde\mu) \ , 
 \end{split} \label{eq:F_contracted}
\end{align}
or, in momentum space:
\begin{align}
 \begin{split}
  \mu^\alpha\mu^\beta\calF_{\alpha\beta}(K;\mu,\tilde\mu) &= i\mu^\alpha\mu^\beta K_\alpha{}^{\dot\alpha}\calA_{\beta\dot\alpha}(K;\mu,\tilde\mu) \ ; \\
  \tilde\mu^{\dot\alpha}\tilde\mu^{\dot\beta}\calF_{\dot\alpha\dot\beta}(K;\mu,\tilde\mu) &= i\tilde\mu^{\dot\alpha}\tilde\mu^{\dot\beta} K^\alpha{}_{\dot\alpha}\calA_{\alpha\dot\beta}(K;\mu,\tilde\mu) \ .
 \end{split} \label{eq:F_contracted_Fourier}
\end{align}
We are now ready to evaluate the Fourier integral \eqref{eq:spinor_modes_final_F} w.r.t. the null time $u$ along the future horizon's lightray. The $u$ dependence in the integral comes from the Fourier factor $e^{i\langle\tilde\mu\mu\rangle u/2}$ in \eqref{eq:spinor_modes_final_F}, from the $u^{-3}$ scaling factor in \eqref{eq:F_components}, and from the $e^{iK\cdot x}$ plane-wave factor in \eqref{eq:A_Fourier}, where $x^\mu$ depends on $u$ as in \eqref{eq:mu_lightray}. The $u$ integral thus takes the form (taking into account also the energy factor $\langle\tilde\mu\mu\rangle^{-2}$ from \eqref{eq:spinor_modes_final_F}):
\begin{align}
 \frac{1}{\langle\tilde\mu\mu\rangle^2}\int_{-\infty}^\infty \frac{du}{u^3}\,e^{i\left(\frac{\langle\tilde\mu\mu\rangle u}{2} - \frac{\langle\mu K\tilde\mu]}{\langle\tilde\mu\mu\rangle u}\right)} = 
  \frac{\sign(\langle\tilde\mu\mu\rangle)}{4}\int_{-\infty}^\infty \frac{dU}{U^3}\,e^{i\left(U - \frac{\langle\mu K\tilde\mu]}{2U}\right)} \ ,
\end{align}
where we rescaled the integration variable as $U\equiv \frac{1}{2}\langle\tilde\mu\mu\rangle u$. We see that the dependence on energy $\langle\tilde\mu\mu\rangle$ -- a non-Lorentz-invariant vestige of the 3d formalism -- has been reduced to a Lorentz-invariant sign. Let's now evaluate the integral by considering it in the complex $U$ plane. At $U\rightarrow\infty$, the integration contour can be closed from above. We must also deform the contour around the essential singularity at $U=0$. The deformation that leads to a well-defined answer is the one for which $\Im(\langle\mu K\tilde\mu]/U)$ is negative. Thus, for $\langle\mu K\tilde\mu]<0$, we must bypass $U=0$ from below. The contour is then equivalent to a circle around $U=0$, and the integral evaluates to the Bessel function of the first kind $J_2$:
\begin{align}
 \oint\frac{dU}{U^3}\,e^{i\left(U - \frac{\langle\mu K\tilde\mu]}{2U}\right)} = \frac{4\pi i}{\langle\mu K\tilde\mu]}\,J_2\!\left(\sqrt{-2\langle\mu K\tilde\mu]} \right) \ . \label{eq:J2}
\end{align}
The result \eqref{eq:J2} is reminiscent of the one found in \cite{Albrychiewicz:2020ruh} for an on-shell spin-0 field. We now turn to the case $\langle\mu K\tilde\mu]>0$. Here, we must bypass $U=0$ from above, resulting in a closed contour with no singularities inside, so the integral vanishes. This makes sense, since $\langle\mu K\tilde\mu]>0$ implies that $K^{\alpha\dot\alpha}$ (a 4-momentum in our flat frame adapted to the past horizon) and $\mu^\alpha\tilde\mu^{\dot\alpha}$ (a 4-momentum in a different flat frame, adapted to the future horizon) have energies of opposite sign with respect to the lightlike coordinate $u$.

Reinstating the polarization factors, we obtain the final data on the future horizon in terms of the non-linear bulk field as:
\begin{align}
 \begin{split}
   c^+_{\text{out}}(\mu,\tilde\mu) &= -2\pi i\sign(\langle\tilde\mu\mu\rangle)\mu^\alpha \mu^\beta \int_{\langle\mu K\tilde\mu] < 0} d^4K\,
     \frac{J_2\!\left(\sqrt{-2\langle\mu K\tilde\mu]} \right)}{\langle\mu K\tilde\mu]}\,K_\alpha{}^{\dot\alpha}\calA_{\beta\dot\alpha}(K;\mu,\tilde\mu) \ ; \\
   c^-_{\text{out}}(\mu,\tilde\mu) &= -2\pi i\sign(\langle\tilde\mu\mu\rangle)\tilde\mu^{\dot\alpha}\tilde\mu^{\dot\beta} \int_{\langle\mu K\tilde\mu] < 0} d^4K\,
     \frac{J_2\!\left(\sqrt{-2\langle\mu K\tilde\mu]} \right)}{\langle\mu K\tilde\mu]}\,K^\alpha{}_{\dot\alpha}\calA_{\alpha\dot\beta}(K;\mu,\tilde\mu) \ ,
 \end{split} \label{eq:final_modes_K_A}
\end{align}
or, in terms of the field strength:
\begin{align}
 \begin{split}
   c^+_{\text{out}}(\mu,\tilde\mu) &= -2\pi\sign(\langle\tilde\mu\mu\rangle)\mu^\alpha \mu^\beta \int_{\langle\mu K\tilde\mu] < 0} d^4K\,
    \frac{J_2\!\left(\sqrt{-2\langle\mu K\tilde\mu]} \right)}{\langle\mu K\tilde\mu]}\,\calF_{\alpha\beta}(K;\mu,\tilde\mu) \ ; \\
   c^-_{\text{out}}(\mu,\tilde\mu) &= -2\pi\sign(\langle\tilde\mu\mu\rangle)\tilde\mu^{\dot\alpha}\tilde\mu^{\dot\beta} \int_{\langle\mu K\tilde\mu] < 0} d^4K\,
    \frac{J_2\!\left(\sqrt{-2\langle\mu K\tilde\mu]} \right)}{\langle\mu K\tilde\mu]}\,\calF_{\dot\alpha\dot\beta}(K;\mu,\tilde\mu) \ .
 \end{split} \label{eq:final_modes_K_F}
\end{align}
Plugging in the field's perturbative expansion \eqref{eq:A_expansion} or \eqref{eq:F_right_expansion}-\eqref{eq:F_left_expansion}, the $d^4K$ integral goes away, because the momentum $K^\mu$ in  \eqref{eq:A_expansion},\eqref{eq:F_right_expansion}-\eqref{eq:F_left_expansion} is always just a sum \eqref{eq:K_sum} of initial momenta $k^\mu$. Thus, our expression for $c^\pm_{\text{out}}$ as functionals of $c^\pm_{\text{in}}$ finally takes the form:
\begin{align}
 \begin{split}
   &c^\pm_{\text{out}}(\mu,\tilde\mu) = 2\pi\sign(\langle\tilde\mu\mu\rangle) \sum_{n=1}^\infty \int_{1\dots n} 
      \frac{\theta(-\langle\mu K_{1\dots n}\tilde\mu])\,J_2\!\left(\sqrt{-2\langle\mu K_{1\dots n}\tilde\mu]} \right)}{\langle\mu K_{1\dots n}\tilde\mu]} \\
   &\quad\times \sum_{(h_1,\dots,h_n)} \calS(1^{h_1},\dots,n^{h_n};\mu,\tilde\mu,\pm)\,
      c^{h_1}_{\text{in}}(\lambda_1,\tilde\lambda_1)\dots c^{h_n}_{\text{in}}(\lambda_i,\tilde\lambda_i) \ ,
 \end{split} \label{eq:kinematics}
\end{align}
where $\theta$ is the step function, and the ``amplitudes'' $\calS$ are related to the perturbative expansions \eqref{eq:A_expansion},\eqref{eq:F_right_expansion}-\eqref{eq:F_left_expansion} of the bulk potential and field strength via:
\begin{align}
 \begin{split}
   \calS(1^{h_1},\dots,n^{h_n};\mu,\tilde\mu,+) &= -i\mu^\alpha\mu^\beta (K_{1\dots n})_\alpha{}^{\dot\alpha}\,a_{\beta\dot\alpha}(1^{h_1},\dots,n^{h_n};\mu,\tilde\mu) \\
     &= -\mu^\alpha\mu^\beta f_{\alpha\beta}(1^{h_1},\dots,n^{h_n};\mu,\tilde\mu) \ ; \\
   \calS(1^{h_1},\dots,n^{h_n};\mu,\tilde\mu,-) 
     &= -i\tilde\mu^{\dot\alpha}\tilde\mu^{\dot\beta} (K_{1\dots n})^\alpha{}_{\dot\alpha}\,a_{\alpha\dot\beta}(1^{h_1},\dots,n^{h_n};\mu,\tilde\mu) \\
     &= -\tilde\mu^{\dot\alpha}\tilde\mu^{\dot\beta} f_{\dot\alpha\dot\beta}(1^{h_1},\dots,n^{h_n};\mu,\tilde\mu) \ .
 \end{split} \label{eq:amplitudes_from_fields}
\end{align}
The overall minus signs in \eqref{eq:kinematics}-\eqref{eq:amplitudes_from_fields} are inserted for later convenience.

We have thus reduced the static-patch scattering problem to a Minkowski-space problem of calculating (certain components of) the non-linear field functional \eqref{eq:A_expansion} or \eqref{eq:F_right_expansion}-\eqref{eq:F_left_expansion}, i.e the color-ordered ``amplitudes'' $a^{\{i,j,\dots\}}_{\alpha\dot\alpha}$ or $(f^{\{i,j,\dots\}}_{\alpha\beta},f^{\{i,j,\dots\}}_{\dot\alpha\dot\beta})$ with a single off-shell leg. In our present formalism, the free-field propagation between the static-patch horizons, first considered in \cite{David:2019mos}, is described by the trivial ``2-point amplitudes'':
\begin{align}
 \begin{split}
   \calS(+;-) &= -[\tilde\lambda\tilde\mu]^2 \ ; \quad \calS(-;+) = -\langle\lambda\mu\rangle^2 \ ; \\
   \calS(+;+) &= \calS(-;-) = 0 \ .
 \end{split} \label{eq:2_point}
\end{align}

\section{N\textsuperscript{-1}MHV scattering} \label{sec:pre_MHV}

In this section, we present the results for tree-level static-patch scattering with N\textsuperscript{-1}MHV helicities, i.e. with \emph{one} of the external leg's helicities negative and the rest positive. As we will see, the N\textsuperscript{-2}MHV amplitudes, with \emph{all} helicities positive, vanish. The N\textsuperscript{-1}MHV amplitudes (and the vanishing of the N\textsuperscript{-2}MHV ones) can be obtained by plugging known classical solutions of Yang-Mills in Minkowski space \cite{Bardeen:1995gk,Rosly:1996vr} into our master kinematical prescription \eqref{eq:kinematics}-\eqref{eq:amplitudes_from_fields}. Here, we review these solutions, adding some clarifications to the original treatments. We begin in section \ref{sec:pre_MHV:self_dual} with a purely self-dual solution; from this, we'll read off the N\textsuperscript{-1}MHV static-patch amplitude in which the negative helicity is on the \emph{outgoing} leg. Then, in section \ref{sec:pre_MHV:f}, we write the linearized left-handed (i.e. anti-self-dual) field strength perturbation over this self-dual solution; from this, we'll read off the N\textsuperscript{-1}MHV static-patch amplitude in which the negative helicity is on one of the \emph{ingoing} legs.

\subsection{Self-dual solution} \label{sec:pre_MHV:self_dual}

In this subsection, we focus on the $c^-_{\text{in}}$-independent piece of the non-linear field \eqref{eq:A_expansion}, i.e. the part of $A_{\alpha\dot\alpha}$ that only depends on the right-handed initial data $c^+_{\text{in}}(\lambda,\tilde\lambda)$. At tree level, this is given by a self-dual solution to the YM field equations, i.e. a solution with purely right-handed field strength, which will generate the N\textsuperscript{-1}MHV static patch amplitudes $\calS(1^+,\dots,n^+;-)$. Let's now describe this self-dual solution, following \cite{Rosly:1996vr}. We will be somewhat less general than the authors of \cite{Rosly:1996vr}, by continuing to work with a constant, i.e. $(\lambda,\tilde\lambda)$-independent, gauge spinor $q^\alpha$. 

The key to the construction of \cite{Rosly:1996vr} is a gauge group element $g(x;q,\rho)$, where $q^\alpha$ and $\rho^\alpha$ are two left-handed spinors ($q^\alpha$ will end up assuming its role as gauge spinor in \eqref{eq:A_expansion}, while $\rho^\alpha$ is a new spinor variable):
\begin{align}
 g(x;q,\rho) = 1 + \langle\rho q\rangle \sum_{n=1}^\infty \int_{1\dots n} \frac{c^+_{\text{in}}(\lambda_1,\tilde\lambda_1)\dots c^+_{\text{in}}(\lambda_n,\tilde\lambda_n)\,e^{iK_{1\dots n}\cdot x}}
   {\langle\rho\lambda_1\rangle\langle\lambda_1\lambda_2\rangle\dots\langle\lambda_{n-1}\lambda_n\rangle\langle\lambda_n q\rangle} \ . \label{eq:g}
\end{align}
Interchanging the spinors $(q,\rho)$ inverts the group element:
\begin{align}
 g^{-1}(x;q,\rho) = g(x;\rho,q) \ . \label{eq:g_inverse}
\end{align}
We can prove this by explicitly writing out all the terms in the product:
\begin{align}
 \begin{split}
   g(x;\rho,q)g(x;q,\rho) = 1 + \langle\rho q\rangle \sum_{n=1}^\infty \int_{1\dots n} &\frac{e^{iK_{1\dots n}\cdot x}\,c^+_{\text{in}}(\lambda_1,\tilde\lambda_1)\dots c^+_{\text{in}}(\lambda_n,\tilde\lambda_n)}
     {\langle q\lambda_1\rangle\langle\lambda_1\lambda_2\rangle\dots\langle\lambda_{n-1}\lambda_n\rangle\langle\lambda_n q\rangle} \\
     &\times\left( \frac{\langle q\lambda_1\rangle}{\langle\rho\lambda_1\rangle} - \frac{\langle\lambda_n q\rangle}{\langle\lambda_n\rho\rangle} 
        + \sum_{i=1}^{n-1}\frac{\langle q\rho\rangle\langle\lambda_i\lambda_{i+1}\rangle}{\langle\lambda_i\rho\rangle\langle\rho\lambda_{i+1}\rangle} \right) \ .
 \end{split} \label{eq:g_inverse_g}
\end{align}
We apply the Schouten identity to the numerators in the sum over $i$:
\begin{align}
  \langle q\rho\rangle\langle\lambda_i\lambda_{i+1}\rangle = \langle q\lambda_i\rangle\langle\rho\lambda_{i+1}\rangle - \langle q\lambda_{i+1}\rangle\langle\rho\lambda_i\rangle \ ,
\end{align} 
which rearranges the corresponding fractions as:
\begin{align}
 \frac{\langle q\rho\rangle\langle\lambda_i\lambda_{i+1}\rangle}{\langle\lambda_i\rho\rangle\langle\rho\lambda_{i+1}\rangle} = \frac{\langle q\lambda_{i+1}\rangle}{\langle\rho\lambda_{i+1}\rangle} - \frac{\langle q\lambda_i\rangle}{\langle\rho\lambda_i\rangle} \ .
 \label{eq:Schouten_frac} 
\end{align}
All the terms in the sum over $n$ in \eqref{eq:g_inverse_g} now cancel, thus proving eq. \eqref{eq:g_inverse}. 

We can now use the group element \eqref{eq:g} to define the self-dual non-linear solution, which we denote by $A^{(0)}_{\alpha\dot\alpha}$:
\begin{align}
 \rho^\alpha g^{-1}(x;q,\rho)\del_{\alpha\dot\alpha}g(x;q,\rho) = \rho^\alpha A^{(0)}_{\alpha\dot\alpha}(x;q) \ . \label{eq:g_A}
\end{align}
The highly non-trivial part of eq. \eqref{eq:g_A} is that it is indeed linear in $\rho^\alpha$, so that $A^{(0)}_{\alpha\dot\alpha}$ does not depend on $\rho^\alpha$. This can be verified by direct computation, analogously to \eqref{eq:g_inverse_g}, with the gradient producing momentum factors as in \eqref{eq:plane_wave}:
\begin{align}
\begin{split}
  \rho^\alpha g^{-1}\del_{\alpha\dot\alpha}g ={}& i\langle\rho q\rangle \sum_{n=1}^\infty \int_{1\dots n} \frac{e^{iK_{1\dots n}\cdot x}\,c^+_{\text{in}}(\lambda_1,\tilde\lambda_1)\dots c^+_{\text{in}}(\lambda_n,\tilde\lambda_n)}
   {\langle q\lambda_1\rangle\langle\lambda_1\lambda_2\rangle\dots\langle\lambda_{n-1}\lambda_n\rangle\langle\lambda_n q\rangle} \\
   &\times\left( \frac{\langle q\lambda_1\rangle}{\langle\rho\lambda_1\rangle}\sum_{j=1}^n \langle\lambda_j\rho\rangle(\tilde\lambda_j)_{\dot\alpha}
    + \sum_{i=1}^{n-1}\frac{\langle q\rho\rangle\langle\lambda_i\lambda_{i+1}\rangle}{\langle\lambda_i\rho\rangle\langle\rho\lambda_{i+1}\rangle}\sum_{j=i+1}^n \langle\lambda_j\rho\rangle(\tilde\lambda_j)_{\dot\alpha} \right) \ .
\end{split}
\end{align}
Applying again the Schouten identity as in \eqref{eq:Schouten_frac}, we find that most of the terms cancel, leaving just a single sum over momenta at each $n$, which rearranges as:
\begin{align}
 \begin{split}
   \rho^\alpha g^{-1}\del_{\alpha\dot\alpha}g &= i\langle\rho q\rangle q^\alpha \sum_{n=1}^\infty \int_{1\dots n} 
      \frac{(K_{1\dots n})_{\alpha\dot\alpha}\,e^{iK_{1\dots n}\cdot x}\,c^+_{\text{in}}(\lambda_1,\tilde\lambda_1)\dots c^+_{\text{in}}(\lambda_n,\tilde\lambda_n)}
      {\langle q\lambda_1\rangle\langle\lambda_1\lambda_2\rangle\dots\langle\lambda_{n-1}\lambda_n\rangle\langle\lambda_n q\rangle} \\
    &= \langle\rho q\rangle q^\alpha\del_{\alpha\dot\alpha}\Phi(x;q) \ ,
 \end{split} \label{eq:g_grad_g}
\end{align}
where we denoted:
\begin{align}
 \Phi(x;q) \equiv \sum_{n=1}^\infty \int_{1\dots n} \frac{e^{iK_{1\dots n}\cdot x}\,c^+_{\text{in}}(\lambda_1,\tilde\lambda_1)\dots c^+_{\text{in}}(\lambda_n,\tilde\lambda_n)}
     {\langle q\lambda_1\rangle\langle\lambda_1\lambda_2\rangle\dots\langle\lambda_{n-1}\lambda_n\rangle\langle\lambda_n q\rangle} = \lim_{\rho^\alpha\rightarrow q^\alpha}\frac{g(x;q,\rho) - 1}{\langle\rho q\rangle} \ .
\end{align}
Eq. \eqref{eq:g_grad_g} is linear in $\rho^\alpha$ as promised, and we can read off the field $A^{(0)}_{\alpha\dot\alpha}$ as:
\begin{align}
 \begin{split}
   A^{(0)}_{\alpha\dot\alpha}(x;q) &= -iq_\alpha q^\beta \sum_{n=1}^\infty \int_{1\dots n} 
      \frac{(K_{1\dots n})_{\beta\dot\alpha}\,e^{iK_{1\dots n}\cdot x}\,c^+_{\text{in}}(\lambda_1,\tilde\lambda_1)\dots c^+_{\text{in}}(\lambda_n,\tilde\lambda_n)}
      {\langle q\lambda_1\rangle\langle\lambda_1\lambda_2\rangle\dots\langle\lambda_{n-1}\lambda_n\rangle\langle\lambda_n q\rangle} \\
      &= -q_\alpha q^\beta\del_{\beta\dot\alpha}\Phi(x;q) \ .
 \end{split} \label{eq:A_self_dual}
\end{align}
The $n=1$ piece of \eqref{eq:A_self_dual} clearly coincides with the right-handed part of the linearized field \eqref{eq:linearized_A},\eqref{eq:linearized_a}. It remains to show that the non-linear corrections make for a self-dual YM solution. This follows from the structure of eq. \eqref{eq:g_A}, which is similar to a flatness condition. In particular, eq. \eqref{eq:g_A} directly implies that the contraction of the left-handed field strength with $\rho^\alpha\rho^\beta$ vanishes:
\begin{align}
 \rho^\alpha\rho^\beta F^{(0)}_{\alpha\beta}(x;q) =  \rho^\alpha\rho^\beta \left(\del_\alpha{}^{\dot\alpha}A^{(0)}_{\beta\dot\alpha}(x;q) + A^{(0)}_\alpha{}^{\dot\alpha}(x;q)A^{(0)}_{\beta\dot\alpha}(x;q) \right) = 0 \ . \label{eq:rho_F}
\end{align}
Since this holds for \emph{any} value of $\rho^\alpha$, we conclude that $F^{(0)}_{\alpha\beta}$ itself vanishes. Thus, $A^{(0)}_{\alpha\dot\alpha}$ describes a self-dual field as promised, and therefore automatically solves the YM field equations. As a corrolary, the N\textsuperscript{-2}MHV amplitudes all vanish:
\begin{align}
 \calS(1^+,\dots,n^+;+) = 0 \ . \label{eq:pre_pre_MHV}
\end{align}
We can now read off from \eqref{eq:A_self_dual} the potential's Taylor coefficients $a_{\alpha\dot\alpha}(1^+,\dots,n^+;q,\tilde q)$:
\begin{align}
 a_{\alpha\dot\alpha}(1^+,\dots,n^+;q,\tilde q) =
   -\frac{iq_\alpha q^\beta (K_{1\dots n})_{\beta\dot\alpha}}{\langle q\lambda_1\rangle\langle\lambda_1\lambda_2\rangle\dots\langle\lambda_{n-1}\lambda_n\rangle\langle\lambda_n q\rangle} \ . \label{eq:a_self_dual}
\end{align}
Substituting $q^\alpha = \mu^\alpha$ and plugging into \eqref{eq:amplitudes_from_fields}, we obtain the N\textsuperscript{-1}MHV static-patch amplitude:
\begin{align}
 \begin{split}
   \calS(1^+,\dots,n^+;\mu,\tilde\mu,-) 
     &= -i\tilde\mu^{\dot\alpha}\tilde\mu^{\dot\beta} (K_{1\dots n})^\alpha{}_{\dot\alpha}\,a_{\alpha\dot\beta}(1^+,\dots,n^+;\mu,\tilde\mu) \\
     &= \frac{\langle\mu K_{1\dots n}\tilde\mu]^2}{\langle\mu\lambda_1\rangle\langle\lambda_1\lambda_2\rangle\dots\langle\lambda_{n-1}\lambda_n\rangle\langle\lambda_n\mu\rangle} \ .
 \end{split} \label{eq:pre_MHV_from_self_dual}
\end{align}
Some further properties of the self-dual solution \eqref{eq:A_self_dual} will be useful below. First, it does not depend on $\tilde q^{\dot\alpha}$. As a result, the gauge condition $\langle qA\tilde q] = 0$ extends into the stronger condition:
\begin{align}
 q^\alpha A^{(0)}_{\alpha\dot\alpha}(x;q) = 0 \ , \label{eq:q_A}
\end{align} 
which is trivial to check. Second, we can evaluate the (purely right-handed) field strength of $A^{(0)}_{\alpha\dot\alpha}$. This is easy, because the piece quadratic in $A^{(0)}_{\alpha\dot\alpha}$ vanishes, so we only get the contribution from the derivative:
\begin{align}
 F^{(0)}_{\dot\alpha\dot\beta}(x;q) = \del^\alpha{}_{\dot\alpha}A^{(0)}_{\alpha\dot\beta}(x;q) = -q^\alpha q^\beta\del_{\alpha\dot\alpha}\del_{\beta\dot\beta}\Phi(x;q) \ .
\end{align}
In terms of Taylor coefficients, this corresponds to:
\begin{align}
 f_{\dot\alpha\dot\beta}(1^+,\dots,n^+;q,\tilde q)
   &= -\frac{q^\alpha q^\beta (K_{1\dots n})_{\alpha\dot\alpha}(K_{1\dots n})_{\beta\dot\beta}}{\langle q\lambda_1\rangle\langle\lambda_1\lambda_2\rangle\dots\langle\lambda_{n-1}\lambda_n\rangle\langle\lambda_n q\rangle} \ ; \label{eq:f_self_dual} \\
 f_{\alpha\beta}(1^+,\dots,n^+;q,\tilde q) &= 0 \ .  
\end{align}
Finally, we can find the gauge transformation that relates the gauge fields $A^{(0)}_{\alpha\dot\alpha}(x;q)$ with different values of $q^\alpha$. Denoting two such values by $(q,q')$, it turns out that the necessary gauge parameter is simply $g(x;q,q')$:
\begin{align}
 A^{(0)}_{\alpha\dot\alpha}(x;q) = g^{-1}(x;q,q')\left(\del_{\alpha\dot\alpha} + A^{(0)}_{\alpha\dot\alpha}(x;q')\right)g(x;q,q') \ . \label{eq:A0_gauge_transformation}
\end{align}
Since the Weyl spinor space is 2-dimensional, it is enough to verify the contractions of this equation with $q^\alpha$ and $q'^\alpha$. These follow directly from eqs. \eqref{eq:g_inverse}, \eqref{eq:g_A} and \eqref{eq:q_A}.

The entire derivation can of course be repeated with opposite chiralities. The anti-self-dual gauge potential reads:
\begin{align}
 \tilde A^{(0)}_{\alpha\dot\alpha}(x;\tilde q) &= -i\tilde q_{\dot\alpha}\tilde q^{\dot\beta} \sum_{n=1}^\infty \int_{1\dots n} 
     \frac{(K_{1\dots n})_{\alpha\dot\beta}\,e^{iK_{1\dots n}\cdot x}\,c^-_{\text{in}}(\lambda_1,\tilde\lambda_1)\dots c^-_{\text{in}}(\lambda_n,\tilde\lambda_n)}
       {[\tilde q\tilde\lambda_1][\tilde\lambda_1\tilde\lambda_2]\dots[\tilde\lambda_{n-1}\tilde\lambda_n][\tilde\lambda_n\tilde q]} \ , \label{eq:A_anti_self_dual}
\end{align}
which corresponds to field coefficients:
\begin{align}
 a_{\alpha\dot\alpha}(1^-,\dots,n^-;q,\tilde q) 
  &= -\frac{i\tilde q_{\dot\alpha}\tilde q^{\dot\beta} (K_{1\dots n})_{\alpha\dot\beta}}{[\tilde q\tilde\lambda_1][\tilde\lambda_1\tilde\lambda_2]\dots[\tilde\lambda_{n-1}\tilde\lambda_n][\tilde\lambda_n\tilde q]} \ ; \\
 f_{\alpha\beta}(1^-,\dots,n^-;q,\tilde q) 
  &= -\frac{\tilde q^{\dot\alpha}\tilde q^{\dot\beta} (K_{1\dots n})_{\alpha\dot\alpha}(K_{1\dots n})_{\beta\dot\beta}}{[\tilde q\tilde\lambda_1][\tilde\lambda_1\tilde\lambda_2]\dots[\tilde\lambda_{n-1}\tilde\lambda_n][\tilde\lambda_n\tilde q]} \ ; \label{eq:f_self_dual_opposite} \\
 f_{\dot\alpha\dot\beta}(1^-,\dots,n^-;q,\tilde q) &= 0 \ .    
\end{align}
and leads to the anti-N\textsuperscript{-1}MHV amplitude with negative helicities on all ingoing legs (with the anti-N\textsuperscript{-2}MHV amplitude vanishing):
\begin{align}
 \calS(1^-,\dots,n^-;+) &= \frac{\langle\mu K_{1\dots n}\tilde\mu]^2}{[\tilde\mu\tilde\lambda_1][\tilde\lambda_1\tilde\lambda_2]\dots[\tilde\lambda_{n-1}\tilde\lambda_n][\tilde\lambda_n\tilde\mu]} \ ; \label{eq:pre_MHV_from_self_dual_opposite} \\
 \calS(1^-,\dots,n^-;-) &= 0 \ .
\end{align}

\subsection{Anti-self-dual field strength perturbation} \label{sec:pre_MHV:f}

Having constructed the (perturbatively) most general self-dual field solution \eqref{eq:A_self_dual}, we now turn to construct a linearized anti-self-dual perturbation over it. It will be sufficient for our purposes to consider just the left-handed field strength $F^{(1)}_{\alpha\beta}$ of this perturbation. By linearity, we can discuss separately the perturbations due to left-handed initial data $c^-_{\text{in}}(\lambda,\tilde\lambda)$ with different values $(\lambda,\tilde\lambda)=(\lambda',\tilde\lambda')$ of the spinor-helicity variables. Thus, we consider a perturbation which, at the non-interacting level, is given simply by:
\begin{align}
 F^{(1)}_{\alpha\beta}(x;q) = \lambda'_\alpha\lambda'_\beta\,e^{ik'\cdot x}\,c^-_{\text{in}}(\lambda',\tilde\lambda') + O(c^+_{\text{in}}) \ . \label{eq:F_1_linear}
\end{align}
Here, the $O(c^+_{\text{in}})$ corrections are due to the interaction with the self-dual background \eqref{eq:A_self_dual}, and $q^\alpha$ sets the gauge in which the latter is defined. This interaction is described by the left-handed half of the YM equations, at first order in the perturbation:
\begin{align}
 \del^\beta{}_{\dot\alpha} F^{(1)}_{\beta\alpha} + [A^{(0)\beta}{}_{\dot\alpha}, F^{(1)}_{\beta\alpha}] = 0 \ . \label{eq:F_1_propagation}
\end{align}
Luckily, there is a gauge in which the interaction becomes trivial. Indeed, in the gauge $q^\alpha=\lambda'^\alpha$, when we plug the linearized solution from \eqref{eq:F_1_linear} into the field equation \eqref{eq:F_1_propagation}, we find that the interaction term vanishes, thanks to \eqref{eq:q_A}. In this gauge, then, the solution is just the non-interacting one. To obtain the result at general $q^\alpha$, we just need to apply the gauge transformation \eqref{eq:A0_gauge_transformation}:
\begin{align}
 F^{(1)}_{\alpha\beta}(x;q) = \lambda'_\alpha\lambda'_\beta\,e^{ik'\cdot x}\, g^{-1}(x;q,\lambda')\,c^-_{\text{in}}(\lambda',\tilde\lambda')\,g(x;q,\lambda') \ . \label{eq:F_1}
\end{align}
Plugging in our expressions \eqref{eq:g}-\eqref{eq:g_inverse} for $g$ and $g^{-1}$, we can read off from \eqref{eq:F_1} the field strength coefficients $f_{\alpha\beta}$ with one ingoing negative helicity on the $i$'th leg:
\begin{align}
 f_{\alpha\beta}(1^+,\dots,i^-,\dots,n^+;q,\tilde q) 
   = -\frac{\langle q\lambda_i\rangle^2(\lambda_i)_\alpha(\lambda_i)_\beta}{\langle q\lambda_1\rangle\langle\lambda_1\lambda_2\rangle\dots\langle\lambda_{n-1}\lambda_n\rangle\langle\lambda_n q\rangle} \ . \label{eq:f_ASD_perturbation}
\end{align}
Plugging into \eqref{eq:amplitudes_from_fields}, we obtain the corresponding N\textsuperscript{-1}MHV static-patch amplitude:
\begin{align}
 \calS(1^+,\dots,i^-,\dots,n^+;+) = \frac{\langle\mu\lambda_i\rangle^4}{\langle\mu\lambda_1\rangle\langle\lambda_1\lambda_2\rangle\dots\langle\lambda_{n-1}\lambda_n\rangle\langle\lambda_n\mu\rangle} \ . \label{eq:pre_MHV_from_f_left}
\end{align}
This is immediately recognizable as the Parke-Taylor formula for MHV scattering \cite{Parke:1986gb} in the context of the Minkowski S-matrix: 
\begin{align}
 \calM(1^+,\dots,i^-,\dots,n^+,(n+1)^-) = \frac{\langle\lambda_{n+1}\lambda_i\rangle^4}{\langle\lambda_{n+1}\lambda_1\rangle\langle\lambda_1\lambda_2\rangle\dots\langle\lambda_{n-1}\lambda_n\rangle\langle\lambda_n\lambda_{n+1}\rangle} \ . \label{eq:Parke_Taylor}
\end{align}
However, the helicities in \eqref{eq:pre_MHV_from_f_left} are N\textsuperscript{-1}MHV, and $\mu_\alpha$ in our context is \emph{not} the spinor square root of the final momentum $K_{1\dots n}^\mu$, which isn't even on-shell. Nevertheless, as we'll see in the next section, the similarity between these amplitudes is not a coincidence.

Once again, the entire analysis can be repeated with the opposite chiralities, yielding the field strength coefficients:
\begin{align}
 f_{\dot\alpha\dot\beta}(1^-,\dots,i^+,\dots,n^-;q,\tilde q) 
   = -\frac{[\tilde q\tilde\lambda_i]^2(\tilde\lambda_i)_{\dot\alpha}(\tilde\lambda_i)_{\dot\beta}}{[\tilde q\tilde\lambda_1][\tilde\lambda_1\tilde\lambda_2]\dots[\tilde\lambda_{n-1}\tilde\lambda_n][\tilde\lambda_n\tilde q]} \ . \label{eq:f_ASD_perturbation_opposite}
\end{align}
and the amplitude:
\begin{align}
 \calS(1^-,\dots,i^+,\dots,n^-;-) = \frac{[\tilde\mu\tilde\lambda_i]^4}{[\tilde\mu\tilde\lambda_1][\tilde\lambda_1\tilde\lambda_2]\dots[\tilde\lambda_{n-1}\tilde\lambda_n][\tilde\lambda_n\tilde\mu]} \ . \label{eq:pre_MHV_from_f_left_opposite} 
\end{align}

\section{Poles, Minkowski S-matrix and BCFW-type recursion} \label{sec:poles}

In this section, we zoom back out from calculating specific amplitudes to discussing general properties of the framework. In particular, we examine the pole behavior of the non-linear tree-level potential $A_{\alpha\dot\alpha}$, its field strength $F_{\alpha\beta},F_{\dot\alpha\beta}$, and the resulting tree-level static-patch amplitudes \eqref{eq:amplitudes_from_fields}. In section \ref{sec:poles:S_matrix}, we discuss how poles in the field strength, which encode the usual tree-level Minkowski S-matrix, are related to \emph{finite components} of the field strength of opposite chirality, and thus to the static-patch amplitudes \eqref{eq:amplitudes_from_fields}. Then, in section \ref{sec:poles:BCFW}-\ref{sec:poles:BCFW_proof}, we define and prove a BCFW recursion relation for the static-patch amplitudes. We will use this BCFW recursion in section \ref{sec:MHV}, to calculate the MHV static-patch amplitude $\calS(1^+,\dots,i^-,\dots,n^+;-)$.

\subsection{Minkowski S-matrix as special case of the static-patch amplitudes} \label{sec:poles:S_matrix}

Intuitively, the usual Minkowski S-matrix should be somehow contained in the static-patch amplitudes \eqref{eq:amplitudes_from_fields}: after all, we can always just send the origin $x^\mu=0$ of our future horizon towards future-timelike Minkowski infinity by an infinitely large time translation. In this section, then, we'll see how exactly the Minkowski S-matrix is related to our static-patch amplitudes $\calS$.

Consider the potential $\calA_{\alpha\dot\alpha}(K;q,\tilde q)$ and field strength $\calF_{\alpha\beta}(K;q,\tilde q),\calF_{\dot\alpha\dot\beta}(K;q,\tilde q)$ in momentum space, in a lightcone gauge $\langle qA\tilde q] = 0$. Specifically ,consider their non-linear parts, of order $n\geq 2$, in the initial data. When the outgoing momentum approaches a lightlike value $K^{\alpha\dot\alpha} = \lambda^\alpha\tilde\lambda^{\dot\alpha}$, the potential and field strength will generally develop $\sim 1/K^2$ poles, whose residues are determined by the usual Minkowski S-matrix (up to our slightly non-standard choice of a retarded $i\varepsilon$ prescription, which ensures exactly one leg to be outgoing, regardless of energy signs). Importantly, this pole at lightlike $K^\mu$ is \emph{not} present in any local product of fields, i.e. in any non-linear term in the field equations. Therefore, the pole's residue satisfies the \emph{linearized} field equations, as expected for an on-shell outgoing particle. Similarly, the residue transforms linearly under gauge transformations, as in Maxwell theory; in particular, the field strength's residue is gauge-invariant. The field strength near $K^{\alpha\dot\alpha} = \lambda^\alpha\tilde\lambda^{\dot\alpha}$ thus takes the form:
\begin{align}
 \begin{split}
   \calF_{\alpha\beta}(K;q,\tilde q) &= \lambda_\alpha\lambda_\beta\left( \frac{\delta(K^2)}{2\pi^2}\,c^-_{\text{in}}(\lambda,\tilde\lambda) + \frac{1}{K^2}\,b^+(\lambda,\tilde\lambda) \right) + \text{finite part} \ ; \\
   \calF_{\dot\alpha\dot\beta}(K;q,\tilde q) &= \tilde\lambda_{\dot\alpha}\tilde\lambda_{\dot\beta}\left( \frac{\delta(K^2)}{2\pi^2}\,c^+_{\text{in}}(\lambda,\tilde\lambda) + \frac{1}{K^2}\,b^-(\lambda,\tilde\lambda) \right) + \text{finite part} \ .
 \end{split} \label{eq:F_poles} 
\end{align}
Here, the first term in the parentheses is the linearized, on-shell field strength, while the second term is the pole as described above. Neither depends on the choice of gauge $(q,\tilde q)$. The residue coefficients $b^\pm(\lambda,\tilde\lambda)$ are functionals of the initial data $c^\pm_{\text{in}}$, whose Taylor coefficients are the usual S-matrix amplitudes (with a minus sign, in our conventions). Explicitly, the Taylor expansion w.r.t. $c^\pm_{\text{in}}$ of the residue term in \eqref{eq:F_poles} takes the form:
\begin{align}
 \begin{split}
   \lim_{K_{1\dots n}^{\gamma\dot\gamma} \rightarrow \lambda^\gamma\tilde\lambda^{\dot\gamma}}K^2 f_{\alpha\beta}(1^{h_1},\dots,n^{h_n};q,\tilde q) &= -\lambda_\alpha\lambda_\beta\,\calM(1^{h_1},\dots,n^{h_n},\{\lambda,-\tilde\lambda,+\}) \ ; \\
   \lim_{K_{1\dots n}^{\gamma\dot\gamma} \rightarrow \lambda^\gamma\tilde\lambda^{\dot\gamma}}K^2 f_{\dot\alpha\dot\beta}(1^{h_1},\dots,n^{h_n};q,\tilde q) 
     &= -\tilde\lambda_{\dot\alpha}\tilde\lambda_{\dot\beta}\,\calM(1^{h_1},\dots,n^{h_n},\{\lambda,-\tilde\lambda,-\}) \ ,
 \end{split} \label{eq:res_f_S}
\end{align}
where $\calM$ denotes an $(n+1)$-point Minkowski S-matrix amplitude. The flipped sign on $\tilde\lambda_{\dot\alpha}$ in its argument simply reverses the final leg's 4-momentum, so as to treat ingoing and outgoing 4-momenta on an equal footing (here, by making them all ingoing). As a general reference on the relationship between tree-level S-matrix amplitudes and classical field solutions, see e.g. \cite{Boulware:1968zz}.

The special case $n=2$ requires separate consideration. There, the square of the outgoing momentum is given by $K^2 = -\langle\lambda_1\lambda_2\rangle[\tilde\lambda_1\tilde\lambda_2]$, and we can discuss separately poles due to $\langle\lambda_1\lambda_2\rangle\rightarrow 0$ and poles due to $[\tilde\lambda_1\tilde\lambda_2]\rightarrow 0$. We already calculated the field strengths in which these poles can arise: these are given by the $n=2$ cases of \eqref{eq:f_self_dual},\eqref{eq:f_self_dual_opposite},\eqref{eq:f_ASD_perturbation},\eqref{eq:f_ASD_perturbation_opposite}. By inspection, we see that \eqref{eq:f_self_dual},\eqref{eq:f_ASD_perturbation} have poles only at $\langle\lambda_1\lambda_2\rangle\rightarrow 0$, while their opposite-chirality counterparts \eqref{eq:f_self_dual_opposite},\eqref{eq:f_ASD_perturbation_opposite} have poles only at $[\tilde\lambda_1\tilde\lambda_2]\rightarrow 0$. This matches the complex kinematics of the Minkowski S-matrix, where we have $\calM(+,+,-)\neq 0$ at $\langle\lambda_1\lambda_2\rangle\rightarrow 0$ but not at $[\tilde\lambda_1\tilde\lambda_2]\rightarrow 0$, and vice versa for $\calM(-,-,+)$.

The non-linear gauge potential $\calA_{\alpha\dot\alpha}$ that corresponds to the field strengths \eqref{eq:F_poles} reads:
\begin{align}
 \begin{split}
   \calA_{\alpha\dot\alpha}(K;q,\tilde q) ={}& -i\frac{\lambda_\alpha\tilde q_{\dot\alpha}}{[\tilde q\tilde\lambda]}\left( \frac{\delta(K^2)}{2\pi^2}\,c^-_{\text{in}}(\lambda,\tilde\lambda) + \frac{1}{K^2}\,b^+(\lambda,\tilde\lambda) \right) \\
     &- i\frac{q_\alpha\tilde\lambda_{\dot\alpha}}{\langle q\lambda\rangle}\left( \frac{\delta(K^2)}{2\pi^2}\,c^+_{\text{in}}(\lambda,\tilde\lambda) + \frac{1}{K^2}\,b^-(\lambda,\tilde\lambda) \right) + \text{finite part} \ .
 \end{split} \label{eq:A_pole}
\end{align}
Now, consider the contractions $\lambda^\alpha\lambda^\beta K_\alpha{}^{\dot\alpha}\calA_{\beta\dot\alpha}$ and $\tilde\lambda^{\dot\alpha}\tilde\lambda^{\dot\beta}K^\alpha{}_{\dot\alpha}\calA_{\alpha\dot\beta}$, of the sort that appear in our static-patch amplitudes \eqref{eq:amplitudes_from_fields}. In the limit $K^{\alpha\dot\alpha} = \lambda^\alpha\tilde\lambda^{\dot\alpha}$, we have $\lambda^\alpha K_\alpha{}^{\dot\alpha} = \tilde\lambda^{\dot\alpha} K^\alpha{}_{\dot\alpha} = 0$; therefore, the contractions can get nonzero contributions only from the pole pieces of \eqref{eq:A_pole}. These are easy to evaluate, using the fact that, near the pole, we can approximate $K^2 \approx \langle\lambda K\tilde\lambda]$. We find:
\begin{align}
\begin{split}
   \lim_{K^{\gamma\dot\gamma} \rightarrow \lambda^\gamma\tilde\lambda^{\dot\gamma}} i\lambda^\alpha\lambda^\beta K_\alpha{}^{\dot\alpha}\calA_{\beta\dot\alpha}(K;q,\tilde q) &= b^-(\lambda,\tilde\lambda) \ ; \\
   \lim_{K^{\gamma\dot\gamma} \rightarrow \lambda^\gamma\tilde\lambda^{\dot\gamma}} i\tilde\lambda^{\dot\alpha}\tilde\lambda^{\dot\beta} K^\alpha{}_{\dot\alpha}\calA_{\alpha\dot\beta}(K;q,\tilde q) &= b^+(\lambda,\tilde\lambda) \ .
 \end{split} \label{eq:F_residues}
\end{align}
In particular, the RHS again does not depend on $(q,\tilde q)$. We can now take the limit $(q,\tilde q)\rightarrow (\lambda,\tilde\lambda)$, in which we recognize the LHS of \eqref{eq:F_residues} as the generating functions for static-patch amplitudes \eqref{eq:amplitudes_from_fields}. Taylor-expanding in $c^\pm_{\text{in}}$, we conclude:
\begin{align}
 \begin{split}
   \lim_{K_{1\dots n}^{\gamma\dot\gamma} \rightarrow \lambda^\gamma\tilde\lambda^{\dot\gamma}} K^2 f_{\alpha\beta}(1^{h_1},\dots,n^{h_n};\lambda,\tilde\lambda) &= -\lambda_\alpha\lambda_\beta \calS(1^{h_1},\dots,n^{h_n};\lambda,\tilde\lambda,-) \ ; \\
   \lim_{K_{1\dots n}^{\gamma\dot\gamma} \rightarrow \lambda^\gamma\tilde\lambda^{\dot\gamma}} K^2 f_{\dot\alpha\dot\beta}(1^{h_1},\dots,n^{h_n};\lambda,\tilde\lambda) 
      &= -\tilde\lambda_{\dot\alpha}\tilde\lambda_{\dot\beta} \calS(1^{h_1},\dots,n^{h_n};\lambda,\tilde\lambda,+) \ ,
 \end{split} \label{eq:res_f_M}
\end{align}
where the amplitudes $\calS$ on the RHS are evaluated at $K_{1\dots n}^{\alpha\dot\alpha} = \lambda^\alpha\tilde\lambda^{\dot\alpha}$. Comparing with \eqref{eq:res_f_S}, we see that the Minkowski S-matrix amplitudes $\calM$ are a special case of our static-patch amplitudes $\calS$, evaluated \emph{at on-shell outgoing momentum} $K_{1\dots n}^{\alpha\dot\alpha} = \lambda^\alpha\tilde\lambda^{\dot\alpha}$, and with an \emph{opposite helicity on the outgoing leg}: 
\begin{align}
 \calM(1^{h_1},\dots,n^{h_n},\{\lambda,\tilde\lambda,h\}) = \calS(1^{h_1},\dots,n^{h_n};\lambda,-\tilde\lambda,-h) \ . \label{eq:S_M}
\end{align}
As a special case, the N\textsuperscript{-1}MHV static-patch amplitude $\calS(1^+,\dots,i^-,\dots,n^+;+)$, when evaluated at $K_{1\dots n}^{\alpha\dot\alpha} = \lambda^\alpha\tilde\lambda^{\dot\alpha}$, should reproduce the Parke-Taylor MHV formula \eqref{eq:Parke_Taylor}. As we've seen in \eqref{eq:pre_MHV_from_f_left}, the two in fact agree for \emph{general} values of $K_{1\dots n}^{\alpha\dot\alpha}$. This stronger agreement is not a coincidence either: as we'll see below, the two amplitudes are governed by essentially the same BCFW recursion relations.

The relation \eqref{eq:S_M} has one apparent exception: it suggests that the 3-point Minkowski S-matrix amplitude $\calM(+,+,-)$ should equal the static patch amplitude $\calS(+,+;+)$; however, the latter is equal to zero, according to \eqref{eq:pre_pre_MHV}. It turns out that this is an order-of-limits ambiguity: if we calculate $\calS(+,+;+)$ within the same limiting procedure as the one that led to \eqref{eq:S_M}, we find a non-zero answer that agrees with $\calM(+,+,-)$ (at necessarily complex momenta, as usual for the 3-point $\calM$ amplitude). Indeed, consider the potential coefficients \eqref{eq:a_self_dual} with $n=2$ ingoing legs:
\begin{align}
 a_{\alpha\dot\alpha}(+,+;q,\tilde q) = -\frac{iq_\alpha q^\beta (K_{12})_{\beta\dot\alpha}}{\langle q\lambda_1\rangle\langle\lambda_1\lambda_2\rangle\langle\lambda_2 q\rangle} \ .
\end{align}
Using $(K_{12})^2 = -\langle\lambda_1\lambda_2\rangle[\tilde\lambda_1\tilde\lambda_2]$, the contraction from \eqref{eq:F_residues} reads:
\begin{align}
 -i\lambda^\alpha\lambda^\beta(K_{12})_\alpha{}^{\dot\alpha} a_{\beta\dot\alpha}(+,+;q,\tilde q) = \frac{[\tilde\lambda_1\tilde\lambda_2]\langle q\lambda\rangle^2}{\langle q\lambda_1\rangle\langle q\lambda_2\rangle} \ . \label{eq:3_point_subtlety_raw}
\end{align}
If we now set $q^\alpha=\lambda^\alpha$, we'll get zero, as in \eqref{eq:pre_pre_MHV}. Instead, let us first take the limit of lightlike $K_{12}^\mu$ via $\langle\lambda_1\lambda_2\rangle\rightarrow 0$. This can be expressed as:
\begin{align}
 \lambda_2^\alpha = w\lambda_1^\alpha \ ; \quad K_{12}^{\alpha\dot\alpha} = \lambda^\alpha\tilde\lambda^{\dot\alpha} \ ; \quad \lambda^\alpha = \lambda_1^\alpha \ ; \quad \tilde\lambda^{\dot\alpha} = \tilde\lambda_1^{\dot\alpha} + w\tilde\lambda_2^{\dot\alpha} \ ,
\end{align}
where $w$ is some scalar. In this limit, eq. \eqref{eq:3_point_subtlety_raw} becomes:
\begin{align}
 -i\lambda^\alpha\lambda^\beta(K_{12})_\alpha{}^{\dot\alpha} a_{\beta\dot\alpha}(+,+;q,\tilde q) = \frac{[\tilde\lambda_1\tilde\lambda_2]}{w} = \frac{[\tilde\lambda_1\tilde\lambda_2]^3}{[\tilde\lambda\tilde\lambda_1][\tilde\lambda_2\tilde\lambda]} \ . \label{eq:3_point_subtlety}
\end{align}
As in \eqref{eq:F_residues}, the RHS is now $(q,\tilde q)$-independent, and we can trivially take the limit $(q,\tilde q)\rightarrow (\lambda,\tilde\lambda)$. We then recognize the two sides of eq. \eqref{eq:3_point_subtlety} as:
\begin{align}
 \calS(1^+,2^+;\lambda,\tilde\lambda,+) = \calM(1^+,2^+,\{\lambda,\tilde\lambda,-\}) \ ,
\end{align}
in agreement with eq. \eqref{eq:S_M} (since the amplitudes in this case are even in $\tilde\lambda^{\dot\alpha}$, flipping its sign has no consequence).

\subsection{BCFW-type recursion} \label{sec:poles:BCFW}

In this section, we define a BCFW-type recursion relation for the tree-level static-patch amplitudes $\calS$. These recursion relations reduce a static-patch amplitude to a Minkowski S-matrix amplitude $\calM$ of the same size, plus products of smaller amplitudes. The recursion can be applied whenever we can find an ingoing leg with the same helicity sign as the outgoing one; this is always the case, \emph{except} for the amplitudes $\calS(1^+,\dots,n^+;-)$ and $\calS(1^-,\dots,n^-;+)$, which we already calculated in section \ref{sec:pre_MHV:self_dual} from the self-dual solution \eqref{eq:A_self_dual}. Thus, our recursion will reduce any amplitude $\calS$ down to:
\begin{enumerate}
	\item The Minkowski S-matrix amplitudes $\calM$ (which in turn can be subjected to the \emph{usual} BCFW recursion).
	\item The static-patch amplitudes $\calS(1^+,\dots,n^+;-)$ and $\calS(1^-,\dots,n^-;+)$ from \eqref{eq:pre_MHV_from_self_dual},\eqref{eq:pre_MHV_from_self_dual_opposite}.
\end{enumerate}
We now proceed to construct the recursion. We single out one of the ingoing legs, e.g. leg number $i$. This, together with the outgoing leg, will form the two external legs involved in the BCFW shift. We assume for concreteness that our singled-out ingoing leg has negative helicity. We then shift its \emph{right}-handed spinor-helicity variable, as:
\begin{align}
 \tilde\lambda_i^{\dot\alpha} \rightarrow \tilde\lambda_i^{\dot\alpha} + z\tilde\mu^{\dot\alpha} \ . \label{eq:shift}
\end{align}
Here, $z$ is a complex variable, while $(\mu,\tilde\mu)$ are the spinor-helicity variables on the future horizon (which, in our Minkowski treatment, are simply defining the lightcone gauge $\langle\mu A\tilde\mu] = 0$). The shift \eqref{eq:shift} changes the 4-momentum of our ingoing leg by $z\lambda_i^\alpha\tilde\mu^{\dot\alpha}$. The same shift then applies to the outgoing 4-momentum $K_{1\dots n}^{\alpha\dot\alpha}$; similarly, it applies to any internal leg that includes the $i$'th one as a summand, i.e. to the sum $K_{j\dots l}^{\alpha\dot\alpha}$ of any consecutive set of ingoing momenta with $j\leq i\leq l$. Now, the static-patch amplitude $\calS$ will have a pole whenever the shift takes one of these momenta on-shell (not counting the $i$'th ingoing momentum itself, which is already on-shell). The key claim is then that the amplitude's original value at $z=0$ can be recovered from the residues at these poles:
\begin{align}
 \calS = -\sum_{\substack{j\leq i\leq l \\ j<l}}\frac{1}{z_{j\dots l}}\,\underset{z=z_{j\dots l}}{\Res}\calS(z) \ . \label{eq:M_from_residues}
\end{align}
This is equivalent to the statement that the contour integral $\oint\frac{\calS(z)}{z}dz$ at infinity vanishes, for which it is sufficient that $\calS(z)$ itself vanishes there:
\begin{align}
 \lim_{z\rightarrow\infty}\calS(z) = 0 \ . \label{eq:vanishing_at_infty}
\end{align}
As we will prove in section \ref{sec:poles:BCFW_proof}, this is indeed the case \emph{if the helicity of the outgoing leg is the same as of the shifted ingoing leg}. For now, let us unpack the content of eq. \eqref{eq:M_from_residues}. The value of $z$ at which the 4-momentum $K_{j\dots l}^\mu$ becomes lightlike reads:
\begin{align}
 z_{j\dots l} = -\frac{K_{j\dots l}^2}{\langle\lambda_i K_{j\dots l}\tilde\mu]} \ .
\end{align}
At this value, the deformed momentum assumes an on-shell value defined by spinors $(\lambda,\tilde\lambda)$ as follows:
\begin{align}
 &K_{j\dots l}^{\alpha\dot\alpha}(z) = K_{j\dots l}^{\alpha\dot\alpha} + z_{j\dots l}\lambda_i^\alpha\tilde\mu^{\dot\alpha} \equiv \lambda^\alpha\tilde\lambda^{\dot\alpha} \ ; \\
 &\lambda^\alpha = K_{j\dots l}^{\alpha\dot\alpha}\tilde\mu_{\dot\alpha} \ ; \quad \tilde\lambda^{\dot\alpha} = -\frac{(\lambda_i)_\alpha K_{j\dots l}^{\alpha\dot\alpha}}{\langle\lambda_i K_{j\dots l}\tilde\mu]} \ , \label{eq:pole_lambda}
\end{align} 
up to the freedom of rescaling $\lambda_\alpha$ and $\tilde\lambda_{\dot\alpha}$ by opposite factors. Our distance from the pole can be parameterized by the contraction:
\begin{align}
 \langle\lambda K_{j\dots l}(z)\tilde\lambda] = (z-z_{j\dots l})\langle\lambda\lambda_i\rangle[\tilde\mu\tilde\lambda] = (z-z_{j\dots l})\langle\lambda_i K_{j\dots l}\tilde\mu] \ .
\end{align}
Near the pole, the field on the newly on-shell leg is described by on-shell plane waves \eqref{eq:F_poles},\eqref{eq:A_pole}, in general with both helicities, whose coefficients are given by Minkowski S-matrix amplitudes, as in \eqref{eq:res_f_S}. These on-shell waves then feed into our overall amplitude $\calS(z)$ as a new ingoing leg, with spinor-helicity variables given by \eqref{eq:pole_lambda}. Putting everything together, we obtain the recursion relation:
\begin{align}
 \begin{split}
   &\calS(1^{h_1},\dots,i^-,\dots,n^{h_n};\mu,\tilde\mu,-) = -\sum_{\substack{j\leq i\leq l \\ j<l}}\frac{1}{K_{j\dots l}^2}\sum_{h = \pm} \\
   &\quad \times \calM\!\left(j^{h_j},\dots,\left\{\lambda_i,\tilde\lambda_i-\frac{K_{j\dots l}^2\tilde\mu}{\langle\lambda_i K_{j\dots l}\tilde\mu]},- \right\},\dots,l^{h_l},
     \left\{K_{j\dots l}^{\alpha\dot\alpha}\tilde\mu_{\dot\alpha},\frac{(\lambda_i)_\alpha K_{j\dots l}^{\alpha\dot\alpha}}{\langle\lambda_i K_{j\dots l}\tilde\mu]},h \right\} \right) \\
   &\quad \times \calS\!\left(1^{h_1},\dots,\left\{K_{j\dots l}^{\alpha\dot\alpha}\tilde\mu_{\dot\alpha},-\frac{(\lambda_i)_\alpha K_{j\dots l}^{\alpha\dot\alpha}}{\langle\lambda_i K_{j\dots l}\tilde\mu]},-h \right\},\dots,n^{h_n};\mu,\tilde\mu,- \right) \ .
 \end{split} \label{eq:BCFW}
\end{align} 
Here, the original static-patch amplitude $\calS$ on the LHS has $n$ ingoing legs and one outgoing, with negative helicities on the $i$'th ingoing leg and on the outgoing one. The Minkowski S-matrix amplitude $\calM$ on the RHS takes a contiguous subset $(j,\dots,l)$ of these ingoing legs (including the $i$'th leg, whose momentum is shifted), and fuses them into an internal on-shell leg described by \eqref{eq:pole_lambda}, with helicity $h$ (which is summed over); the remaining static-patch amplitude $\calS$ accepts this new on-shell leg in place of the $(j,\dots,l)$ subset. Most of the terms in \eqref{eq:BCFW} contain amplitudes with strictly fewer external legs than the original one on the LHS. The one exception is the term with $(j,l)=(1,n)$, which as an $\calM$ amplitude with the same number of legs (and the same helicities) as the original $\calS$ on the LHS, times a trivial 2-point $\calS$ ``amplitude'' from \eqref{eq:2_point}. 

\begin{figure}%
	\centering%
	\includegraphics[scale=1]{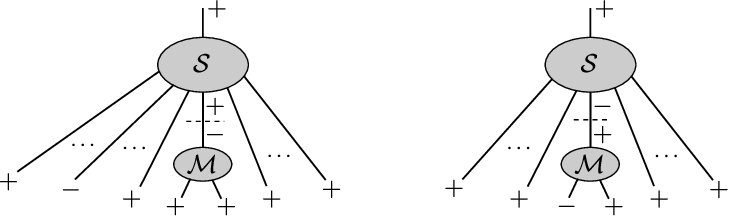} \\
	\caption{The two types of pole contributions in the recursion of the N\textsuperscript{-1}MHV static-patch amplitudes $\calS(1^+,\dots,i^-,\dots,n^+;+)$ via eq. \eqref{eq:BCFW_opposite}. In both cases, the amplitude reduces to a smaller one of the same type, times the 3-point Minkowski S-matrix amplitude $\calM(+,+,-)$.}
	\label{fig:Recursion1} 
\end{figure}

The analogous recursion formula with \emph{positive} helicities on the outgoing leg and on the shifted ingoing one reads:
\begin{align}
 \begin{split}
   &\calS(1^{h_1},\dots,i^+,\dots,n^{h_n};\mu,\tilde\mu,+) = -\sum_{\substack{j\leq i\leq l \\ j<l}}\frac{1}{K_{j\dots l}^2}\sum_{h = \pm} \\
   &\quad \times \calM\!\left(j^{h_j},\dots,\left\{\lambda_i - \frac{K_{j\dots l}^2\mu}{\langle\mu K_{j\dots l}\tilde\lambda_i]},\tilde\lambda_i,+\right \},\dots,l^{h_l},
     \left\{\frac{K_{j\dots l}^{\alpha\dot\alpha}(\tilde\lambda_i)_{\dot\alpha}}{\langle\mu K_{j\dots l}\tilde\lambda_i]},\mu_\alpha K_{j\dots l}^{\alpha\dot\alpha},h \right\} \right) \\
   &\quad \times \calS\!\left(1^{h_1},\dots,\left\{\frac{K_{j\dots l}^{\alpha\dot\alpha}(\tilde\lambda_i)_{\dot\alpha}}{\langle\mu K_{j\dots l}\tilde\lambda_i]},-\mu_\alpha K_{j\dots l}^{\alpha\dot\alpha},-h \right\},\dots,n^{h_n};\mu,\tilde\mu,+ \right) \ .
 \end{split} \label{eq:BCFW_opposite}
\end{align}
As a consistency check, it's easy to verify that the N\textsuperscript{-1}MHV amplitude $\calS(1^+,\dots,i^-,\dots,n^+;+)$ from \eqref{eq:pre_MHV_from_f_left} satisfies the recursion relation \eqref{eq:BCFW_opposite}, while its N\textsuperscript{-1}MHV counterpart $\calS(1^-,\dots,i^+,\dots,n^-;-)$ from \eqref{eq:pre_MHV_from_f_left_opposite} satisfies \eqref{eq:BCFW}. In a slight notational clash, the recursion in these cases can be applied to any ingoing leg \emph{other than} the $i$'th one. Then, depending on whether we chose the first or last ingoing leg $(1,n)$, or an intermediate one, the sum will includes one or two poles. These poles reduce the $(n+1)$-point static-patch amplitude $\calS$ into an $n$-point amplitude of the same type, times a 3-point Minkowski S-matrix amplitude $\calM$ (see Figure \ref{fig:Recursion1}). This is completely analogous to how the usual BCFW recursion works on the MHV and anti-MHV $\calM$ amplitudes. This explains the ``coincidence'' between the static-patch amplitudes \eqref{eq:pre_MHV_from_f_left},\eqref{eq:pre_MHV_from_f_left_opposite} and the Parke-Taylor formula \eqref{eq:Parke_Taylor}.

\subsection{Comparison with scalar field theory} \label{sec:poles:scalar}

This is a good place to draw a comparison with scalar field theories. Consider a scalar theory that is conformal at tree-level, i.e. a conformally massless scalar with $\varphi^4$ interaction. For such a theory, we can pose the static-patch scattering problem, and work out its kinematics, just like in section \ref{sec:kinematics}, but without gauge choices or polarization factors. In particular, eqs. \eqref{eq:kinematics}-\eqref{eq:amplitudes_from_fields} carry through: we just need to remove all helicity signs, all references to the $(\mu,\tilde\mu)$-dependent lightcone gauge, the $\langle\mu K_{1\dots n}\tilde\mu]$ denominator in \eqref{eq:kinematics}, and the $\mu,\tilde\mu,K_{1\dots n}$ prefactors in \eqref{eq:amplitudes_from_fields}. In this way, the problem of static-patch scattering for a conformal scalar theory can be reduced to that of calculating a non-linear bulk field in Minkowski space, as a functional of ingoing linearized plane waves. If we wish, we can of course also consider this Minkowski problem for more general, non-conformal scalar theories; however, we will then lose the original connection with static-patch scattering. 

Consider, then a scalar theory in Minkowski. Unlike Yang-Mills, we know that its Minkowski S-matrix is \emph{not} subject to BCFW recursion, because it doesn't vanish at $z\rightarrow\infty$. On the other hand, our particular recursion statement from section \ref{sec:poles:BCFW} \emph{does} hold for scalar theories. First, it is definitely true that the (off-shell) scalar bulk field can be reduced to its (fully on-shell) S-matrix amplitudes: the two are just related by the amputation of the final $1/K^2$ propagator. Therefore, the static-patch amplitudes for a scalar theory are directly reducible to Minkowski S-matrix amplitudes, even without going through a BCFW-like argument; the only problem is that the scalar Minkowski S-matrix itself is not as well-behaved as in the Yang-Mills case. 

Furthermore, while it isn't \emph{necessary} in the scalar case, the analog of the particular BCFW-type logic from section \ref{sec:poles:BCFW} holds here as well, and is easy to prove. Indeed, let us shift the momentum of an ingoing leg as in \eqref{eq:shift}. This shifts the momentum of every leg that includes this ingoing leg as a summand, via:
\begin{align}
 K_{\alpha\dot\alpha}\rightarrow K_{\alpha\dot\alpha} + z\lambda'_\alpha\tilde\mu_{\dot\alpha} \ . \label{eq:shift_K}
\end{align}
At large $z$, the magnitude-squared of this shifted momentum behaves as:
\begin{align} 
 K^2\rightarrow -z\langle\lambda' K\tilde\mu] \ . \label{eq:shift_K_squared}
\end{align}
Thus, every $1/K^2$ propagator that's affected by the shift will introduce a factor of $1/z$ into the bulk field, i.e. into the static-patch amplitude. And there will always be at least one such propagator, i.e. the one on the outgoing leg (unlike with the Minkowski S-matrix amplitudes, where the outgoing propagator is amputated). Thus, the scalar static-patch amplitude vanishes at least as $\sim 1/z$ at large $z$, as required for the BCFW-like recursion.

\subsection{Proof of the recursion for Yang-Mills theory} \label{sec:poles:BCFW_proof}

We now return to the Yang-Mills case. Again, we want to demonstrate the vanishing \eqref{eq:vanishing_at_infty} of static-patch amplitudes at $z\rightarrow\infty$, which will ensure the validity of the recursion formulas \eqref{eq:BCFW}-\eqref{eq:BCFW_opposite}. In light of section \ref{sec:poles:scalar}, this means our task is to show that YM theory behaves \emph{no worse} than scalar theory in this regard. As usual, our liability will be the extra momentum factors in the YM Lagrangian. As we will show, they can be rendered harmless by careful use of gauge symmetry.

We focus on the case of \eqref{eq:BCFW}, i.e. negative helicity on the shifted ingoing leg. We'll take a similar approach to that in section \ref{sec:pre_MHV}: we will consider the non-linear field that describes the amplitudes as the sum $A^{(0)}_{\alpha\dot\alpha} + A^{(1)}_{\alpha\dot\alpha}$ of a background field and a linearized perturbation. Unlike in section \ref{sec:pre_MHV}, the background field need not be self-dual: it is simply the field composed of all the ingoing legs \emph{other than} the shifted one. The perturbation $A^{(1)}_{\alpha\dot\alpha}$ then describes the shifted ingoing leg, and its propagation through the $A^{(0)}_{\alpha\dot\alpha}$ background. This can itself be organized into a perturbation series. Thus, we write:
\begin{align}
 A^{(1)}_{\alpha\dot\alpha}(x) = \sum_{m=0}^\infty A^{(1;m)}_{\alpha\dot\alpha}(x) \ , \label{eq:A_1_series}
\end{align}
where $A^{(1;0)}_{\alpha\dot\alpha}(x)$ is the non-interacting approximation, and $A^{(1;m)}_{\alpha\dot\alpha}$ with $m>0$ is the correction due to diagrams with $m$ interactions with the background $A^{(0)}_{\alpha\dot\alpha}$. As in section \ref{sec:pre_MHV:f}, we focus on the linearized perturbation at a particular value $(\lambda,\tilde\lambda)=(\lambda',\tilde\lambda')$ of spinor-helicity variables on the shifted ingoing leg. Thus, the non-interacting term in the series \eqref{eq:A_1_series} reads:
\begin{align}
 A^{(1;0)}_{\alpha\dot\alpha}(x) = -i\frac{\lambda'_\alpha \tilde\mu_{\dot\alpha}}{[\tilde\mu\tilde\lambda']}\,e^{ik'\cdot x} \ , \label{eq:A_1_linearized}
\end{align}
where we used $\tilde\mu^{\dot\alpha}$ to fix the gauge-dependent part of the polarization. Our proof of \eqref{eq:vanishing_at_infty} will now consist of two steps:
\begin{enumerate}
	\item We will show that, for $A^{(0)}_{\alpha\dot\alpha}$ in the complexified lightcone gauge $\langle\lambda'A^{(0)}\tilde\mu] = 0$, there \emph{exists} a gauge (not necessarily a lightcone gauge) in which the corrections $A^{(1;m)}_{\alpha\dot\alpha}$ for all $m>0$ vanish at $z\rightarrow\infty$ as $A^{(1;m)}_{\alpha\dot\alpha}\sim 1/z$. 
	\item We will transform into the gauge $\langle\mu A\tilde\mu\rangle = 0$. There, we will show that the contraction $\tilde\mu^{\dot\alpha}\tilde\mu^{\dot\beta} \del^\alpha{}_{\dot\alpha}A^{(1;m)}_{\alpha\dot\beta}$ which generates the static-patch amplitudes $\calS(z)$ again vanishes as $\sim 1/z$, even though $A^{(1;m)}_{\alpha\dot\alpha}$ itself may not.
\end{enumerate}
Let's begin with the first step. The non-interacting term \eqref{eq:A_1_linearized} remains unchanged under the BCFW shift \eqref{eq:shift} (apart from the change to the momentum itself). Let us now study the interacting corrections, by considering their origin in the Yang-Mills field equation. At each order $m>0$, the equation (in momentum space) takes the general form:
\begin{align}
 K^2\calA^{(1;m)}_\mu(K) - K_\mu K^\nu\calA^{(1;m)}_\nu(K) =  \calJ^{(1;m)}_\mu(K) \ ,
\end{align}
with solution:
\begin{align}
 \calA^{(1;m)}_\mu(K) = \frac{1}{K^2}\left(\calJ^{(1;m)}_\mu(K) + \theta^{(1;m)}(K)K_\mu \right) \ . \label{eq:perturbative_solution}
\end{align}
Here, the ``current'' $\calJ^{(1;m)}_\mu$ denotes the interaction terms, which contain exactly one factor of $\calA^{(1;m-1)}_\mu$, some factors of $\calA^{(0)}_\mu$ (either one or two), and at most one momentum factor. The gauge-algebra-valued function $\theta^{(1;m)}(K)$ is arbitrary, and encodes the solution's gauge freedom at each order. 

In our present formalism, the BCFW shift \eqref{eq:shift} consists in shifting the momentum of $A^{(1;m)}_{\alpha\dot\alpha}$ at every order $m$, as in \eqref{eq:shift_K}. At large $z$, this implies the $\sim z$ behavior \eqref{eq:shift_K_squared} for $K^2$. Therefore, the solution \eqref{eq:perturbative_solution} vanishes as $\sim 1/z$ (like in the case of scalar field field theory), \emph{if} the expression in parentheses does not grow with $z$. This can be arranged by suitably tuning $\theta^{(1;m)}(K)$, so long as $\calJ_{\alpha\dot\alpha}^{(1;m)}$ grows with $z$ at most linearly, and only along the shift vector $\lambda'_\alpha\tilde\mu_{\dot\alpha}$. Let us now show that this is indeed the case, assuming that the background field is given in the gauge $\langle\lambda' A^{(0)}\tilde\mu] = 0$. First, note that if $\calA^{(1;m)}_\mu$ at some order $m$ vanishes as $\sim 1/z$, then $\calJ^{(1;m+1)}_\mu$ doesn't grow with $z$ at all. This is because positive powers of $z$ can only arise from factors of the shifted momentum \eqref{eq:shift_K}, but there's at most one such factor in every term in $\calJ^{(1;m+1)}_\mu$, and this factor of $z$ will be canceled by the $\sim 1/z$ behavior of $\calA^{(1;m)}_\mu$. Thus, if $\calA^{(1;m)}_\mu \sim 1/z$, then, with the choice $\theta^{(1;m+1)}=0$, we get $\calA^{(1;m+1)}_\mu \sim 1/z$ at the next order as well. It remains to show that the \emph{first} interacting correction $\calA^{(1;1)}_\mu$ vanishes as $1/z$. Since $\calA^{(1;0)}_{\alpha\dot\alpha}$ is $z$-independent, there is a danger of positive powers of $z$ from the terms in $\calJ^{(1;1)}_\mu$ that contain a factor of the shifted momentum \eqref{eq:shift_K}, or, equivalently, a spacetime gradient acting on $A^{(1;1)}_\mu$. There are three such terms:
\begin{align}
 \calJ^{(1;1)}_\mu(K) = i\int d^4K'\!&\left( 2K'^\nu\!\left[\calA^{(0)}_\nu(K-K'), \calA^{(1;0)}_\mu(K') \right] - K'^\nu\!\left[\calA^{(0)}_\mu(K-K'), \calA^{(1;0)}_\nu(K') \right] \right. \nonumber \\
    &\quad \left. -  K'_\mu\!\left[\calA^{(0)\nu}(K-K'), \calA^{(1;0)}_\nu(K') \right] \right) + \dots \ , \label{eq:J_danger}
\end{align}
where the dots denote terms factors of the shifted momentum. Let's now examine the terms one by one. The first term does not grow with $z$, thanks to our assumed gauge condition $\langle\lambda' A^{(0)}\tilde\mu] = 0$ on the background field. Similarly, the second term doesn't grow with $z$, thanks to the property $\langle\lambda' A^{(1;0)}\tilde\mu] = 0$ of the non-interacting perturbation \eqref{eq:A_1_linearized}. Finally, the third term in \eqref{eq:J_danger} \emph{does} grow linearly with $z$, but only along the shift vector $\lambda'_\alpha\tilde\mu_{\dot\alpha}$, which means that the growth can be canceled by tuning the gauge function $\theta^{(1;1)}$. This completes our proof that, \emph{in a certain gauge}, the corrections $A^{(1;m)}_\mu$ at all orders $m>0$ vanish as $1/z$.

Let us now transform into the gauge $\langle\mu A\tilde\mu] = 0$ that is relevant for the static-patch amplitudes. First, we apply a gauge transformation $g^{(0)}$ that brings the background field $A^{(0)}_{\alpha\dot\alpha}$ from the gauge $\langle\lambda' A^{(0)}\tilde\mu] = 0$ into the desired one $\langle\mu A^{(0)}\tilde\mu] = 0$. This transformation is $z$-independent. It transforms the field perturbation as $A^{(1)}_{\alpha\dot\alpha}\rightarrow (g^{(0)})^{-1}A^{(1)}_{\alpha\dot\alpha}\,g^{(0)}$; this affects only the interacting corrections $A^{(1;m)}_{\alpha\dot\alpha}$ with $m>0$, and does not change their $\sim 1/z$ behavior. What's missing now is a linearized gauge transformation $1 + G^{(1)}$ that would ensure the vanishing of $\langle\mu A^{(1)}\tilde\mu]$. The effect of such a linearized transformation is:
\begin{align}
 A^{(1)}_{\alpha\dot\alpha} \rightarrow A^{(1)}_{\alpha\dot\alpha} + \del_{\alpha\dot\alpha} G^{(1)} + [A^{(0)}_{\alpha\dot\alpha},G^{(1)}] \ . \label{eq:G_transformation}
\end{align}
The $[A^{(0)}_{\alpha\dot\alpha},G^{(1)}]$ term does not affect the $\langle\mu A\tilde\mu] = 0$ gauge, since $A^{(0)}_{\alpha\dot\alpha}$ is already in it. The gradient term can then bring about $\langle\mu A^{(1)}\tilde\mu] = 0$, by choosing $G^{(1)}$ as:
\begin{align}
 G^{(1)}(x) = \int d^4K\,\calG^{(1)}(K)\,e^{iK\cdot x} \ ; \quad \calG^{(1)}(K) = \frac{i\langle\mu\calA^{(1)}(K)\tilde\mu]}{\langle\mu K\tilde\mu]} \ . \label{eq:G}
\end{align}
The non-interacting perturbation \eqref{eq:A_1_linearized} doesn't contribute to \eqref{eq:G}, since it already satisfies $\langle\mu A^{(1;0)}\tilde\mu] = 0$. Therefore, $G^{(1)}$ is proportional to the interacting corrections $A^{(1;m)}_{\alpha\dot\alpha}$ with $m>0$, and thus vanishes at large $z$ as $\sim 1/z$. Coming back to the transformed potential \eqref{eq:G_transformation}, we see that it vanishes as $\sim 1/z$, except for the non-interacting piece \eqref{eq:A_1_linearized} as before, and except for the gradient term $\del_{\alpha\dot\alpha} G^{(1)}$, which, under the shift \eqref{eq:shift_K} develops a $O(z^0)$ piece along $\lambda'_\alpha\tilde\mu_{\dot\alpha}$. All in all, then, in the gauge $\langle\mu A\tilde\mu] = 0$, the perturbation field $A^{(1)}_{\alpha\dot\alpha}$ does not quite vanish at large $z$, but its non-vanishing part is along $\lambda'_\alpha\tilde\mu_{\dot\alpha}$. Therefore, the contraction $\tilde\mu^{\dot\alpha}\tilde\mu^{\dot\beta}\del^\alpha{}_{\dot\alpha}A^{(1)}_{\alpha\dot\beta}$ that generates static-patch amplitudes $\calS$ with negative helicity on the outgoing leg \emph{does} vanish at large $z$. This concludes our proof of the BCFW recursion \eqref{eq:BCFW} with negative helicities on the outgoing leg and on the shifted ingoing one. The proof of the relation \eqref{eq:BCFW_opposite} with both helicities positive is analogous.

\section{MHV scattering} \label{sec:MHV}

\begin{figure}%
	\centering%
	\includegraphics[scale=1]{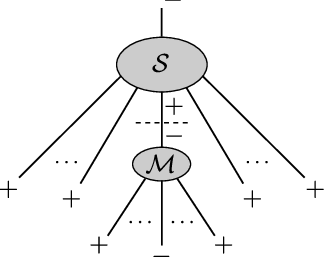} \\
	\caption{A pole contribution in the recursion of the MHV static-patch amplitude $\calS(1^+,\dots,i^-,\dots,n^+;-)$ via eq. \eqref{eq:BCFW}. The amplitude decomposes into a Minkowski S-matrix MHV amplitude, times a simpler static-patch amplitude of the type $\calS(+,\dots,+;-)$.}
	\label{fig:Recursion2} 
\end{figure}

In this section, we apply the recursion formula \eqref{eq:BCFW} to calculate the MHV static-patch amplitudes $\calS(1^+,\dots,i^-,\dots,n^+;-)$. In this simple case, the BCFW shift can only be applied to the $i$'th leg, since it's the only one with the same helicity as the ougoing leg. The recursion involves only a single step, which decomposes $\calS(1^+,\dots,i^-,\dots,n^+;-)$ into products of a Parke-Taylor MHV amplitude $\calM(+,\dots,-,\dots,+,-)$ and a static-patch N\textsuperscript{-1}MHV amplitude $\calS(+,\dots,+;-)$ (see Figure \ref{fig:Recursion2}):
\begin{align}
 \begin{split}
   &\calS(1^+,\dots,i^-,\dots,n^+;\mu,\tilde\mu,-) = -\sum_{\substack{j\leq i\leq l \\ j<l}}\frac{1}{K_{j\dots l}^2} \\
   &\quad \times \calM\!\left(j^+,\dots,\left\{\lambda_i,\tilde\lambda_i-\frac{K_{j\dots l}^2\tilde\mu}{\langle\lambda_i K_{j\dots l}\tilde\mu]},- \right\},\dots,l^+,
    \left\{K_{j\dots l}^{\alpha\dot\alpha}\tilde\mu_{\dot\alpha},\frac{(\lambda_i)_\alpha K_{j\dots l}^{\alpha\dot\alpha}}{\langle\lambda_i K_{j\dots l}\tilde\mu]},- \right\} \right) \\
   &\quad \times \calS\!\left(1^+,\dots,\left\{K_{j\dots l}^{\alpha\dot\alpha}\tilde\mu_{\dot\alpha},-\frac{(\lambda_i)_\alpha K_{j\dots l}^{\alpha\dot\alpha}}{\langle\lambda_i K_{j\dots l}\tilde\mu]},+ \right\},\dots,n^+;\mu,\tilde\mu,- \right) \ .
 \end{split} \label{eq:MHV_BCFW}
\end{align} 
The recursion terminates after this single step, because, unlike what we saw for the N\textsuperscript{-1}MHV case \eqref{eq:pre_MHV_from_f_left},\eqref{eq:pre_MHV_from_f_left_opposite}, it does \emph{not} involve products of smaller MHV amplitudes $\calS(+,\dots,-,\dots,+;-)$ with 3-point amplitudes $\calM(-,+,+)$. This is because the non-vanishing amplitudes $\calM(-,+,+)$ can only be reached by shifting $\lambda_i^\alpha$, not $\tilde\lambda_i^{\dot\alpha}$. 

The partial amplitudes in \eqref{eq:MHV_BCFW} can be evaluated immediately, using eqs. \eqref{eq:Parke_Taylor},\eqref{eq:pre_MHV_from_self_dual}. The MHV static-patch amplitude \eqref{eq:MHV_BCFW} then evaluates to:
\begin{align}
 \begin{split}
   &\calS(1^+,\dots,i^-,\dots,n^+;\mu,\tilde\mu,-) = -\frac{\langle\mu K_{1\dots n}\tilde\mu]^2}{\langle\mu\lambda_1\rangle\langle\lambda_1\lambda_2\rangle\dots\langle\lambda_{n-1}\lambda_n\rangle\langle\lambda_n\mu\rangle} \\
   &\quad \times \sum_{\substack{j\leq i\leq l \\ j<l}} \frac{\langle\lambda_{j-1}\lambda_j\rangle\langle\lambda_l\lambda_{l+1}\rangle\langle\lambda_i K_{j\dots l}\tilde\mu]^4}
     {K_{j\dots l}^2\langle\lambda_{j-1} K_{j\dots l}\tilde\mu]\langle\lambda_j K_{j\dots l}\tilde\mu]\langle\lambda_l K_{j\dots l}\tilde\mu]\langle\lambda_{l+1} K_{j\dots l}\tilde\mu]} \ ,
 \end{split} \label{eq:MHV_result}
\end{align}
where, in the edge cases $j=1$ and $l=n$, one should replace $\lambda_0$ and $\lambda_{n+1}$ respectively by $\mu$. Note that the prefactor before the sum in \eqref{eq:MHV_result} is just the N\textsuperscript{-1}MHV amplitude \eqref{eq:pre_MHV_from_self_dual}. As a consistency check, it's easy to verify that for $n=2$, the ``MHV'' amplitude \eqref{eq:MHV_result} agrees with ``anti-N\textsuperscript{-1}MHV'' amplitude \eqref{eq:pre_MHV_from_f_left_opposite}. Finally, we can of course reverse all helicities in \eqref{eq:MHV_result}, obtaining the anti-MHV static-patch amplitude:
\begin{align}
 \begin{split}
   &\calS(1^-,\dots,i^+,\dots,n^-;\mu,\tilde\mu,+) = -\frac{\langle\mu K_{1\dots n}\tilde\mu]^2}{[\tilde\mu\tilde\lambda_1][\tilde\lambda_1\tilde\lambda_2]\dots[\tilde\lambda_{n-1}\tilde\lambda_n][\tilde\lambda_n\tilde\mu]} \\
   &\quad \times \sum_{\substack{j\leq i\leq l \\ j<l}} \frac{[\tilde\lambda_{j-1}\tilde\lambda_j][\tilde\lambda_l\tilde\lambda_{l+1}]\langle\mu K_{j\dots l}\tilde\lambda_i]^4}
   {K_{j\dots l}^2\langle\mu K_{j\dots l}\tilde\lambda_{j-1}]\langle\mu K_{j\dots l}\tilde\lambda_j]\langle\mu K_{j\dots l}\tilde\lambda_l]\langle\mu K_{j\dots l}\tilde\lambda_{l+1}]} \ ,
 \end{split} \label{eq:MHV_result_opposite}
\end{align}
where any occurrences of $\tilde\lambda_0$ and/or $\tilde\lambda_{n+1}$ should be replaced by $\tilde\mu$.

As is generally the case in BCFW recursion, the individual summands in \eqref{eq:MHV_result}-\eqref{eq:MHV_result_opposite} contain spurious poles, which correspond to an internal propagator going on-shell in the BCFW-shifted amplitude, but not in the original one. These poles are contained in the denominator factors of the form $\langle\lambda K\tilde\mu]$ and $\langle\mu K\tilde\lambda]$. The unphysical poles should of course cancel in the overall amplitude, once the sum is performed. 

\section{Discussion} \label{sec:discuss}

In this paper, we studied tree-level scattering for Yang-Mills theory in a conformally flat causal diamond. Since the theory is conformal, one can consider it in various conformal frames. For us, the most conceptually important causal diamond is the static patch of de Sitter space, hence the terminology ``static-patch amplitudes''. On the other hand, as we've seen, the most convenient conformal frame to actually work in is one where the tips of the causal diamond are at the origin and at (past timelike) infinity of Minkowski space. In this frame, we've demonstrated that the static-patch amplitudes are only slightly more complicated than the usual Minkowski S-matrix, which they include as a limit. This is in contrast with the case of Yang-Mills (A)dS boundary correlators, which appear to be substantially more difficult. The main qualitative difference between our static-patch amplitudes and the Minkowski S-matrix is that they are nonzero at the N\textsuperscript{-1}MHV level. As we have shown, all other amplitudes can be recursively reduced to these N\textsuperscript{-1}MHV ones, ``dressed'' with the Minkowski S-matrix. We applied this recursion to the simplest non-trivial case, calculating the MHV amplitudes $\calS(1^+,\dots,i^-,\dots,n^+;-)$. With some more work, one can of course apply the recursion to more complicated cases, including with the other class of MHV amplitudes $\calS(1^+,\dots,i^-,\dots,j^-,\dots,n^+;+)$, which we did not calculate here.

An interesting open question would be to what extent our techniques can be extended beyond tree-level. For a theory that's conformal at the quantum level, such as $\calN=4$ SYM, this should be relatively straightforward. Without supersymmetry, though, the conformal symmetry of YM theory is broken by loop corrections. Perhaps in perturbation theory, it's somehow possible to treat these violations systematically, and still apply Minkowski methods to static-patch scattering?

Another obvious question is what about perturbative GR in the de Sitter static patch. On one hand, gravity is not conformal even at tree level, and its perturbation theory in de Sitter space is rather painful. On the other hand, it is tempting to speculate that our static-patch results for Yang-Mills, such as the simple formulas for (N\textsuperscript{-1})MHV scattering, or the validity of BCFW recursion, should somehow ``square'' into true statements about GR, along the lines of color-kinematics duality \cite{Bern:2008qj,Armstrong:2020woi,Albayrak:2020fyp}. Going yet further up in spin, it would be interesting to see if any insights from the Yang-Mills case may carry over to higher-spin gravity in de Sitter space. In particular, perhaps the N\textsuperscript{-1}MHV amplitudes \eqref{eq:pre_MHV_from_self_dual}, which are a product of self-dual Yang-Mills theory, can be uplifted into chiral higher-spin theory \cite{Metsaev:1991nb,Metsaev:1991mt,Ponomarev:2016lrm,Skvortsov:2018jea,Skvortsov:2020wtf}.

\section*{Acknowledgements}

We are grateful to Olga Gelfond, Sudip Ghosh, Slava Lysov and Mikhail Vasiliev for many discussions on the topics presented in this paper. This work was supported by the Quantum Gravity Unit of the Okinawa Institute of Science and Technology Graduate University (OIST), which hosted EA as an intern during the project's early stages.

\end{document}